\documentstyle[amssymb,12pt]{article}
 
\begin{document}

\noindent \mbox{\hspace*{63ex}} OCU-PHYS-171, 1999 

\vspace*{10mm}

\begin{center}
{\Large {\bf Out of Equilibrium Quantum Field Theory} \\ 
--- Perturbation Theory and Generalized Boltzmann Equation} 
\end{center}

\hspace*{3ex}

\hspace*{3ex}

\hspace*{3ex}

\begin{center} 
{\large {\sc A. Ni\'{e}gawa}\footnote{ 
E-mail: niegawa@sci.osaka-cu.ac.jp}

{\normalsize\em Department of Physics, Osaka City University } \\ 
{\normalsize\em Sumiyoshi-ku, Osaka 558-8585, Japan} } \\
\end{center} 

\hspace*{2ex}

\hspace*{2ex}

\hspace*{2ex}

\hspace*{2ex}
\begin{center} 
{\large {\bf Abstract}} \\ 
\end{center} 
\begin{quotation} 
This paper describes perturbative framework, on the basis of 
the closed-time-path formalism, in terms of quasi\-particle 
picture for studying quasi\-uniform relativistic quantum field 
systems near equilibrium and non\-equi\-lib\-ri\-um 
quasi\-stationary 
systems. Two calculational schemes are introduced, the one is 
formulated on the basis of the initial-particle distribution 
function and the one is formulated on the basis of the \lq\lq 
physical''-particle distribution function. It is shown that both 
schemes are equivalent and lead to a generalized kinetic or 
Boltzmann equation. Concrete procedure of computing a generic 
amplitude is presented. 

\hspace*{1ex} 
\end{quotation}
\newpage
\setcounter{equation}{0}
\setcounter{section}{0}
\def\theequation{\mbox{\arabic{section}.\arabic{equation}}}
\setcounter{equation}{0}
\setcounter{section}{0}
\def\theequation{\mbox{\arabic{section}.\arabic{equation}}}
\section{Introduction}
Since mid-fifties, efforts have been made to incorporate quantum 
field theory with nonequilibrium statistical 
mechanics\cite{sch,chou,lan,ume}. Necessity of this incorporation 
originated from the field of solid-state physics. Since then, rapid 
progress of the studies of the early universe and the quark-gluon 
plasma, which is expected to be produced in heavy-ion collisions and 
to have existed in the early universe, have further activated this 
field of research (cf., e.g., Refs.~5) and 6)). 

The standard formalism of nonequilibrium statistical quantum-field 
theory is broadly classified into two frameworks, the one is the 
closed-time-path (CTP) formalism \cite{sch,chou,lan,hu,Mor} and the 
one is nonequilibrium thermo field dynamics.\footnote{The 
nonequilibrium thermo field dynamics is initiated by Arimitsu and 
Umezawa\cite{ari} and has developed into sophisticated 
form.\cite{ume,ume1}} In this paper, we employ the former. 
Throughout this paper, we are interested in quasi\-uniform systems 
near equilibrium or non\-equilibrium quasistationary systems, which 
we simply refer to as out-of-equi\-lib\-ri\-um systems. A brief 
discussion is made in \S5 on the relation with thermo field 
dynamics. 

The out-of-equilibrium systems are characterized by two different 
spacetime scales: microscopic or quantum-field-theoretical and 
macroscopic or statistical. The first scale, the 
microscopic-correlation scale, characterizes the range of radiative 
correction to reactions taking place in the system while the second 
scale measures the relaxation of the system. For a weak-coupling 
theory, in which we are interested in this paper, the former scale 
is much smaller than the latter scale. A well-known intuitive 
picture (cf., e.g., Ref.~7) for dealing with such systems is to 
separate spacetime into many \lq\lq cells'' whose characteristic 
size, $L^\mu$ $(\mu = 0, ..., 3)$, is in between the microscopic and 
macroscopic scales. It is assumed that the correlation between 
different cells is negligible in the sense that microscopic or 
elementary reactions can be regarded as taking place in a single 
cell. On the other hand, in a single cell, relaxation phenomena are 
negligible. 

Above intuitive picture may be implemented as follows. Let $\Delta 
(x, y)$ be a generic propagator. For an out-of-equilibrium system, 
$\Delta (x, y)$, with $x - y$ fixed, does not change appreciably in 
the region $|X^\mu - 
X_0^\mu| \lesssim L^\mu$, where $X^\mu \equiv (x^\mu + y^\mu) / 2$ 
is the mid-point and $X_0^\mu$ is an arbitrary spacetime point. 
A self-energy part $\Sigma (x, y)$ enjoys similar property. 
Thus, $X^\mu$ may be used as a label for the spacetime cells and 
is called the macroscopic spacetime coordinates. On the other hand, 
relative spacetime coordinates $x^\mu - y^\mu$ are responsible for 
describing microscopic reactions taking place in a single spacetime 
cell. An inverse Fourier transformation with respect to the relative 
coordinates $x - y$ yields 
\begin{equation} 
\Delta (X; P) \equiv \int d^{\, 4} (x - y)  \, e^{i P \cdot (x - 
y)} \, \Delta (x, y) 
\label{Fourier} 
\end{equation} 
(with $P^\mu = (p^0, {\bf p})$) together with a similar formula for 
$\Sigma$. Above observation shows that, for $|P^\mu| \gtrsim 1 / 
L^\mu$, $P^\mu$ in (\ref{Fourier}) is regarded as the momentum of 
the quasi\-particle participating in the microscopic reaction under 
consideration. 

Microscopic reactions discussed above cause change in the density 
matrix that characterizes the ensemble of the system, through which 
the number densities of quasi\-particles change with macroscopic 
spacetime $X^\mu$. Dealing with this is the subject of the \lq\lq 
next stage,'' where (weak) $X^\mu$-dependence of $\Delta (X; P)$ and 
$\Sigma (X; P)$ are explicitly taken into account. 

The purpose of this paper is to clarify the structure of 
perturbative out-of-equi\-lib\-ri\-um quantum-field theory through 
introducing two mutually equivalent calculational schemes. The one 
(bare-$N$ scheme) is formulated in terms of the initial- or \lq\lq 
bare''-number-density function (\S2~-~\S3) and the one 
(physical-$N$ scheme) is formulated in terms of the \lq\lq 
renormalized''-number-density function (\S4). We demonstrate the 
equivalence of both schemes through deriving a generalized Boltzmann 
equation. Concrete procedure of computing a generic amplitude is 
presented in terms of Feynman rules supplemented with the derivative 
expansion and the generalized Boltzmann equation (\S4). The bare-$N$ 
scheme is the scheme which is directly deduced from first 
principles. To the best knowledge of the present author, this scheme 
has never been comprehensively dealt with. The physical-$N$ scheme 
is the scheme which is employed in the literature. Within the CTP 
formalism, however, no self-consistent deduction of the scheme is 
available. We deduce this scheme from first principles in an 
unambiguous manner. Comparison with related work is made in \S5. 

Throughout this paper, for the sake of definiteness, we take up a 
self-interacting complex-scalar quantum field theory. Generalization 
to other theories is straightforward. 
\section{Closed-time-path formalism} 
To make the paper self-contained, in this section, we give a brief 
review of the CTP formalism. 
\subsection{Preliminaries} 
The Lagrangian (density) reads 
\begin{equation} 
{\cal L} = - \phi^\dagger (\partial^2 + m^2) \phi - 
\frac{\lambda}{4} (\phi^\dagger \phi)^2 . 
\label{Lag} 
\end{equation} 
The CTP formalism is constructed through introducing an oriented 
closed-time path $C$ $(= C_1 \oplus C_2)$ in a complex-time plane, 
which goes from $- \infty$ to $+ \infty$ ($C_1$) and then returns 
back from $+ \infty$ to $- \infty$ ($C_2$). The time argument $x_0$ 
of the fields, $\phi (x)$, is on the time path $C$. The perturbative 
scheme for computing time-path-ordered Green functions is 
summarized as a set of Feynman rules (cf.~Ref. 2)). 

Throughout this paper, we assume, for simplicity,  that the density 
matrix $\rho$ commutes with the charge operator $Q$: $[\rho, Q] = 
0$. (Generalization to the case $[\rho, Q] \neq 0$ is 
straightforward.) Building blocks of Feynman rules are the 
propagator, the vertex and the so-called initial-correlations. The 
propagator $\Delta$ is defined by 
\begin{equation} 
\Delta (x, y) = - i \, \mbox{Tr} \left[ T_C \left( \phi (x) 
\phi^\dagger (y) \right) \, \rho \right] \, \left( \equiv - i 
\langle T_C \left( \phi (x) \phi^\dagger (y) \right) \rangle \right) 
\label{prde} 
\end{equation} 
and the initial correlations are 
\begin{equation} 
{\cal C}_{2 n} \equiv i (-)^n \mbox{Tr} \left[ \, : \phi (x_1) \, 
... \, \phi (x_n) \phi^\dagger (y_1) \, ... \, \phi^\dagger (y_n): 
\, \rho \, \right] \;\;\;\;\;\; (n \geq 2) . 
\label{shoki} 
\end{equation} 
Here $T_C$ is the time-ordering operator along the time-path $C$ and 
$: ... :$ indicates to take normal ordering with respect to the 
creation and annihilation operators in vacuum theory. Throughout 
this paper, we do not deal with ${\cal C}_{2 n}$ $(n \geq 2)$. 
(In Appendix A, we discuss a condition, under which 
${\cal C}_{2 n}$ $(n \geq 2)$ may be ignored when compared to the $2 
n$-point Green function.) 

A field $\phi (x_0, {\bf x})$ with $x_0 \in C_1$ 
$[C_2]$ is called a type-1 [type-2] field, which we write $\phi_1 
(x)$ [$\phi_2 (x)$]. The type-1 field is also called a physical 
field. A classical contour action may be written in the form, 
\begin{eqnarray} 
\int_C d x_0 \int d^{\, 3} x {\cal L} (\phi (x), \phi^\dagger (x)) 
& = & \int_{- \infty}^{+ \infty} d x_0 \int d^{\, 3} x \left[ 
{\cal L} (\phi_1 (x), \phi^\dagger_1 (x)) - {\cal L} (\phi_2 (x), 
\phi^\dagger_2 (x)) \right] \nonumber \\ 
& \equiv & \int d^{\, 4} x \, \hat{\cal L} (x) . 
\label{hat} 
\end{eqnarray} 
Following Refs.~4), 10) and 11), we call $\hat{\cal L} (x)$ the 
hat-Lagrangian (density). One can choose any value for the time $x_0 
= T_{in}$, when interaction-picture fields and Heisenberg-picture 
ones coincide. As in, e.g., Refs.~2) and 3), we choose $T_{in} = - 
\infty$, which we call the initial time. It is well known that the 
theory is ultraviolet (UV) renormalizable. As a matter of fact, in 
general, introduction of renormalization counter terms in vacuum 
theory is sufficient for the out-of-equi\-lib\-ri\-um theory to be 
renormalized. Thus, in the following, we do not write down 
explicitly the UV-re\-normali\-za\-tion counter terms. The free 
hat-Lagrangian $\hat{\cal L}_0$, being bilinear in the fields, is 
obtained from (\ref{Lag}) and (\ref{hat}): 
\begin{equation} 
\hat{\cal L}_0 = - \hat{\phi}^\dagger ( \partial^2 + m^2) 
\hat{\tau_3} \hat{\phi} , 
\label{free-hat} 
\end{equation} 
where $\hat{\tau}_3$ is the third Pauli matrix and 
\[ 
\hat{\phi}^\dagger = \left( \phi_1^\dagger, \, \phi_2^\dagger 
\right) , \;\;\;\;\;\;\; \hat{\phi} = \left( 
\begin{array}{c} 
\phi_1 \\ 
\phi_2 
\end{array} 
\right) . 
\] 
Then, we are naturally lead to a two-component theory: The 
propagator, the vertex and the self-energy part take $2 \times 2$ 
matrix form. We use $\hat{B}$ to denote the $2 \times 2$ matrix 
whose $(i, j)$ component is $B_{i j}$. From (\ref{hat}), we see 
that the vertex matrix $\hat{V}$ has simple structure, $\hat{V} = 
\mbox{diag} (v, - v)$ with $v$ the vertex factor in vacuum theory. 
The propagator (\ref{prde}) is translated into the propagator 
matrix: 
\begin{equation} 
\hat{\Delta} (x, y) = - i \langle T_C \left( \hat{\phi} (x) 
\hat{\phi}^\dagger (y) \right) \rangle , 
\label{denp} 
\end{equation} 
where $T_C$ acts as rearranging the fields in time-path order 
by recalling the above definition of the type-1 and type-2 fields. 
In this and the next sections, taking $\hat{\cal L}_0$ (Eq. 
(\ref{free-hat})) as the free hat-Lagrangian, we construct a 
perturbative scheme, which we call the bare-$N$(umber) scheme. 

Fourier transforming the propagator $\hat{\Delta} (x, y)$ [the 
self-energy part $\hat{\Sigma} (x, y)$] on $x - y$, we have 
$\hat{\Delta} (X; P)$ [$\hat{\Sigma} (X; P)$] (cf. 
Eq.~(\ref{Fourier})). From the setup in \S1, $\hat{\Delta} 
(X; P)$ and $\hat{\Sigma} (X; P)$ vary slowly in $X$. Then, we 
employ the derivative expansion: 
\begin{equation} 
\hat{G} (X; P) = \left[ \, 1 + (X - Y)^\mu \partial_{Y^\mu} 
+ \frac{1}{2} (X - Y)^\mu (X - Y)^\nu \partial_{Y^\mu} 
\partial_{Y^\nu} + ... \right] \hat{G} (Y; P) , 
\label{hosi} 
\end{equation} 
where $\hat{G}$ stands for $\hat{\Delta}$ or $\hat{\Sigma}$. 
Throughout this paper, unless otherwise stated, we keep only first two 
terms in (\ref{hosi}) (gradient approximation):   
\[ 
\hat{G} (X; P) \simeq \hat{G} (Y; P) + (X - Y) \cdot \partial_Y 
\hat{G} (Y; P) , 
\] 
where and in the following, \lq $\simeq$' is used to denote the 
gradient approximation. 
\subsection{Quasiparticle representation for the propagator} 
In this subsection, pigeonholing the standard 
approach,\cite{chou,ume,hu,Mor,ume1} we introduce a quasiparticle 
representation for the propagator. 
\subsubsection{Preliminary} 
From (\ref{free-hat}) and (\ref{denp}), we have 
\[ 
(\partial_x^2 + m^2) \hat{\Delta} (x, y) = (\partial_y^2 + m^2) 
\hat{\Delta} (x, y) = - \hat{\tau}_3 \, \delta^{\, 4} (x - y) 
\] 
and from (\ref{denp}), we can readily obtain\footnote{Note that 
$[\phi (x), \; \phi^\dagger (y)]$ is a c-number function.} 
\begin{eqnarray} 
&& \Delta_{1 1} + \Delta_{2 2} = \Delta_{1 2} + \Delta_{2 1} , 
\label{wa} \\ 
&& \Delta_R = \Delta_{1 1} - \Delta_{1 2} = - i \theta (x_0 - y_0) 
\, \left[ \phi (x), \; \phi^\dagger (y) \right] , 
\label{reta} \\ 
&& \Delta_A = \Delta_{1 1} - \Delta_{2 1} = - i \theta (y_0 - x_0) 
\, \left[ \phi (x), \; \phi^\dagger (y) \right] , 
\label{adva} \\ 
&& \Delta_c = \Delta_{1 2} + \Delta_{2 1} = - i \, \langle \phi (x) 
\phi^\dagger (y) + \phi^\dagger (y) \phi (x) \rangle , 
\label{rac} 
\end{eqnarray} 
where $\Delta_R$, $\Delta_A$ and $\Delta_c$ are retarded, advanced 
and correlation functions, respectively. Equation (\ref{wa}) tells 
us that out of four elements $\Delta_{i j}$ $(i, j = 1, 2)$, only 
three are independent. For independent quantities, we choose 
$\Delta_R$, $\Delta_A$ and $\Delta_c$, which satisfy 
\begin{eqnarray} 
(\partial_x^2 + m^2) \Delta_{R (A)} (x, y) & = & (\partial_y^2 + 
m^2) \Delta_{R (A)} (x, y) = - \delta^{\, 4} (x - y) , \nonumber 
\\ 
(\partial_x^2 + m^2) \Delta_c (x, y) & = & (\partial_y^2 + m^2) 
\Delta_c (x, y) = 0 . 
\label{eq1} 
\end{eqnarray} 
It is to be noted that (\ref{wa}) is valid also for self-energy-part 
resummed propagators (see Appendix B). 
\subsubsection{Quasiparticle representation} 
Now we introduce a short-hand notation $F \cdot G$, which is the 
function whose \lq\lq $(x, y)$ component'' is 
\[ 
[F \cdot G] (x, y) = \int d^{\, 4} z F (x, z) G (z, y) . 
\] 
Motivated by equilibrium thermal field theory 
(ETFT),\cite{lan,ume,umt,eco} we introduce new 
functions\cite{chou,ume,ume1} 
\[ 
f_B^{(\pm)} (x, y) \equiv \delta (x_0 - y_0) \, g_B^{(\pm)} ({\bf 
x}, {\bf y}; x_0) 
\] 
(with \lq $B$' for short of Bare-$N$ scheme) and write $\Delta_c (x, 
y)$ $(\equiv \Delta_c^{(+)} + \Delta_c^{(-)})$ in the form, 
\begin{equation} 
\Delta_c^{(\pm)} \equiv \Delta_R^{(\pm)} \cdot (1 + 2 f_B^{(\pm)}) - 
(1 + 2 f_B^{(\pm)}) \cdot \Delta_A^{(\pm)} , 
\label{Gc} 
\end{equation} 
where $\Delta^{(+)}$'s [$\Delta^{(-)}$'s] stand for the positive 
[negative] frequency part of $\Delta$'s (cf. (\ref{Fourier})) and 
\lq $1$' is the function whose \lq\lq $(x, y)$ component'' is 
$\delta^{\, 4} (x - y)$. (The representation of $\Delta_c$ in 
terms of a vacuum-theory kit is given in Appendix C.) As in ETFT, 
in the case where $\rho$ is diagonal in momentum space, 
$\hat{\Delta} (x, y) = \hat{\Delta} (x - y)$ and 
\[ 
\int d^{\, 3} x \, e^{- i \tau {\bf p} \cdot {\bf x}} g^{(\tau)}_B 
({\bf x}) = \tau \tilde{n}_\tau ({\bf p}) - \theta (- \tau) 
\;\;\;\;\;\; (\tau = \pm) 
\] 
with $\tilde{n}_+$ the number density of the quasiparticle and 
$\tilde{n}_-$ the anti-quasiparticle (cf. Appendix C). In the 
present 
out-of-equilibrium case, $g_B^{(\pm)} ({\bf x}, {\bf y}; x_0)$ 
depends on ${\bf X}$ $(= ({\bf x} + {\bf y} ) / 2)$ and on $x_0$ 
only weakly. Then, in what follows, for $g_B^{(\pm)} ({\bf x}, {\bf 
y}; x_0)$, we use the gradient approximation not only for ${\bf X}$ 
($= ({\bf x} + {\bf y}) / 2$) but also for $x_0$. 

It is worth noting that $\hat{\Delta}$ may be written as (cf. 
(\ref{wa})~-~(\ref{rac}) and (\ref{Gc})) 
\begin{eqnarray} 
\hat{\Delta} & = & \sum_{\tau = \pm} \hat{B}_L (f_B^{(\tau)}) \cdot 
\hat{\Delta}_{diag}^{(\tau)} \cdot \hat{B}_R (f_B^{(\tau)}) , 
\label{aha} \\ 
\hat{\Delta}_{diag} & = & \sum_{\tau = \pm} 
\hat{\Delta}_{diag}^{(\tau)} = \mbox{diag} \left( \Delta_R , \, - 
\Delta_A \right) , \nonumber \\ 
\hat{B}_L (f_B^{(\tau)}) & = & \left( 
\begin{array}{cc} 
1 & \; f_B^{(\tau)} \\ 
1 & \; 1 + f_B^{(\tau)} 
\end{array} 
\right) , \;\;\; \hat{B}_R (f_B^{(\tau)}) = \left( 
\begin{array}{cc} 
1 + f_B^{(\tau)} & \; f_B^{(\tau)} \\ 
1 & \; 1 
\end{array} 
\right) . 
\label{mat} 
\end{eqnarray} 
This representation may be interpreted in terms of \lq\lq 
retarded/advanced-quasi\-par\-ti\-cle picture.'' We introduce \lq\lq 
retarded/advanced-quasi\-par\-ti\-cle fields,'' $\hat{\phi}_{R A}$ 
and $\hat{\phi}^\dagger_{R A}$, through transformations (with 
obvious notation), 
\begin{equation} 
\hat{\phi} = \sum_{\tau = \pm} \hat{\phi}^{(\tau)} = \sum_{\tau 
= \pm}  \hat{B}_L (f_B^{(\tau)}) \cdot \hat{\phi}_{R A}^{(\tau)} , 
\;\;\; \hat{\phi}^\dagger = \sum_{\tau = \pm} \hat{\phi}^{(\tau) 
\dagger} = \sum_{\tau = \pm} \hat{\phi}_{R A}^{(\tau) 
\dagger} \cdot \hat{B}_R (f_B^{(\tau)}) . 
\label{qu-2} 
\end{equation} 
From (\ref{mat}), we see that 
\[ 
\hat{B}_L (f_B^{(\tau)}) \cdot 1 \hat{\tau}_3 \cdot \hat{B}_R 
(f_B^{(\tau)}) = 1 \, \hat{\tau}_3 , 
\] 
which guarantees the canonical commutation relation to be 
preserved: 
\begin{eqnarray*} 
\left[ \hat{\phi} (x), \, \dot{\hat{\phi}}^\dagger (y) \right] 
\delta (x_0 - y_0) & = & \left[ \hat{\phi}_{R A} (x), \, 
\dot{\hat{\phi}}^\dagger_{R A} (y) \right] \delta (x_0 - y_0) 
\nonumber \\ 
& = & i \hat{\tau}_3 \, \delta^4 (x - y) . 
\end{eqnarray*} 
Thus, the ETFT counterpart of (\ref{qu-2}) is called the (thermal) 
Bogoliubov transformation. \cite{umt} The statistical average of 
$T_C \left( \hat{\phi}_{R A} \hat{\phi}^\dagger_{R A} \right)$ (cf. 
(\ref{denp})) assumes the form 
\begin{equation} 
\langle \hat{T}_C \left( \hat{\phi}_{R A} (x) \, 
\hat{\phi}^\dagger_{R A} (y) \right) \rangle = i \hat{\Delta}_{daig} 
(x - y) . 
\label{qu-1} 
\end{equation} 
Thus the fields $\hat{\phi}_{R A}$ and $\hat{\phi}^\dagger_{R A}$ 
describe well-defined propagating modes in the system. It is to be 
noted that $\hat{\phi}^\dagger_{R A}$ is not the hermitian-conjugate 
of $\hat{\phi}_{R A}$, which is a characteristic feature of the 
theory of this type\cite{ume,ume1}. Substituting (\ref{qu-2}) and 
(\ref{qu-1}) into (\ref{denp}), (\ref{wa})~-~(\ref{rac}) with 
(\ref{Gc}) are reproduced. 

Incidentally, nonequilibrium thermo field dynamics\cite{ume,ume1} 
is formulated by taking (\ref{qu-2}) as one of the starting 
hypothesis. 

Now let us go to the $P$-space (cf. (\ref{Fourier})). 
$\Delta_R$ and $\Delta_A$ go to (cf. (\ref{reta}) and (\ref{adva})) 
\begin{equation} 
\Delta_{R (A)} (P) = \frac{1}{P^2 - m^2 \pm i \epsilon (p_0) 0^+} . 
\label{Delta-ra} 
\end{equation} 
It is straightforward to show that (\ref{Gc}) goes to 
\begin{eqnarray} 
\Delta_c (X; P) & \simeq & - 2 \pi i \epsilon (p_0) [1 + 2 
g_B^{(\epsilon (p_0))} (X; {\bf p})] \, \delta (P^2 - m^2) 
\nonumber \\ 
&& + 2 i \partial_{X^\mu} g_B^{(\epsilon (p_0))} (X; {\bf p}) \, 
\partial_{P_\mu} \frac{\bf P}{P^2 - m^2} , 
\label{Deltac} 
\end{eqnarray} 
where 
\[ 
g_B^{(\pm)} (X; {\bf p}) = \int d ({\bf x} - {\bf y}) e^{ - i 
{\bf p} \cdot ({\bf x} - {\bf y})} g^{(\pm)}_B ({\bf x}, {\bf y} ; 
X_0) 
\] 
with ${\bf X} = ({\bf x} + {\bf y}) / 2$. One can easily show that 
(\ref{eq1}) with (\ref{Gc}) yields 
\begin{equation} 
\partial_{X_0} g_B^{(\pm)} (X; {\bf p}) = {\bf p} \cdot \nabla_{\bf 
X} \, g_B^{(\pm)} (X; {\bf p}) = 0 . 
\label{const-b} 
\end{equation} 
Then, the last term in (\ref{Deltac}) vanishes:\footnote{As has been 
mentioned in \S1, (\ref{Deltac-1}) is meaningful in the 
region $|P^\mu| \gtrsim 1 / L^\mu$, where (\ref{Gc}) makes sense. 
(In the region $|P^\mu| \lesssim 1 / L^\mu$, $\Delta_c^{(\pm)}$ in 
(\ref{Gc}) \lq\lq mixes'' with $\Delta^{(\mp)}_{R (A)}$). For 
dealing with the region $|P^\mu| \lesssim 1 / L^\mu$, one should 
return back to the fundamental form (\ref{rac}) or rather to 
(\ref{A1}) in Appendix C. Nevertheless, in most 
practical cases, one can use the form (\ref{Deltac-1}) for whole 
range of $P^\mu$. This can be seen as follows. Let ${\cal T}$ be a 
typical scale(s) of the system under consideration. In the case of 
thermal-equilibrium system, ${\cal T}$ is the temperature of the 
system. Due to 
interactions, an effective mass $M_{ind} (X)$ is induced. In the 
case of $m >> \sqrt{\lambda} {\cal T}$, $M_{ind} (X)$ is not much 
different from $m$ and, for $m \lesssim \sqrt{\lambda} {\cal T}$, a 
tadpole diagram induces $M_{ind} (X)$ of $O (\sqrt{\lambda} {\cal 
T})$. [Since $\sqrt{\lambda} {\cal T}$ (or even $\lambda {\cal T}$) 
is the scale that characterizes microscopic reactions, the setup in 
\S1 shows that $1 / L^\mu << \sqrt{\lambda} {\cal T}$ (or $\lambda 
{\cal T}$).] Since most amplitudes, when computed in perturbation 
theory, are insensitive to the region $|P^\mu| << M_{ind} (X)$ of 
$\Delta_c (X; P)$, the precise form of $\Delta_c (X; P)$ in this 
region is irrelevant. In the following, we keep this in mind. 
Incidentally, in the case of equilibrium thermal QED or QCD $(m = 
0)$, there are some quantities that diverge at infrared limits to 
leading order of hard-thermal-loop resummation scheme\cite{pis,eco}. 
For analyzing such quantities, one should use the fundamental 
formula (\ref{rac}) in the region $|P^\mu| \lesssim 1 / L^\mu$.} 
\begin{equation} 
\Delta_c (X; P) \simeq - 2 \pi i \epsilon (p_0) [1 + 2 
g_B^{(\epsilon (p_0))} (X; {\bf p})] \, \delta (P^2 - m^2) . 
\label{Deltac-1} 
\end{equation} 
\subsubsection{Particle-number density} 
In order to find out the physical meaning\footnote{See also Appendix 
C.} of $g_B^{(\pm)}$, let us compute the \lq\lq free'' or 
nonperturbative part of the current density: 
\[ 
j^\mu (x) \equiv \frac{i}{2} \left[ \phi^\dagger (x) 
\stackrel{\leftrightarrow}{\partial^\mu} \phi (x) - \phi (x) 
\stackrel{\leftrightarrow}{\partial^\mu} \phi^\dagger (x) 
\right] , 
\] 
where $\stackrel{\leftrightarrow}{\partial} \equiv \partial - 
\stackrel{\leftarrow}{\partial}$. Taking the statistical average of 
these quantities, we obtain 
\begin{equation} 
\langle j^\mu (x) \rangle = \frac{1}{2} (\partial_{y_\mu} - 
\partial_{x_\mu}) \Delta_c (x, y) \, \rule[-3mm]{.14mm}{8.5mm} 
\raisebox{-2.85mm}{\scriptsize{$\; x = y$}} . 
\label{cur-1} 
\end{equation} 
Straightforward manipulation using (\ref{Deltac-1}) yields 
\begin{eqnarray} 
\langle j^0 (x) \rangle - \langle 0 | j^0 (x) | 0 \rangle 
& = & \int \frac{d^{\, 3} p}{(2 \pi)^3} 
\, \left[ g^{(+)}_B (x; {\bf p}) + \{ 1 + g^{(-)}_B (x; {\bf p}) 
\} \right] , \nonumber \\ 
\langle {\bf j} (x) \rangle - \langle 0 | {\bf j} (x) \ 0 \rangle 
& = & \int \frac{d^{\, 3} p}{(2 
\pi)^3} \, \frac{\bf p}{E_p} \left[ g^{(+)}_B (x; {\bf p}) - \{ 1 + 
g^{(-)}_B (x; {\bf p}) \} \right] , 
\label{curr1} 
\end{eqnarray} 
where $E_p = \sqrt{p^2 + m^2}$. Equation (\ref{curr1}) is to be 
compared with 
\begin{equation} 
\langle j^\mu (x) \rangle - \langle 0 | j^\mu (x) | 0 \rangle 
= \int \frac{d^{\, 3} p}{(2 \pi)^3} v_B^\mu \left[ N_B^{(+)} (x; 
{\bf p}) - N^{(-)}_B (x; {\bf p}) \} \right] , 
\label{iik} 
\end{equation} 
where $v_B^\mu = (1, {\bf p} / E_p)$ is the four-velocity and 
$N_B^{(+)} (x; {\bf p})$ [$N_B^{(-)} (x; {\bf p})$] is the number 
density (function) of the quasiparticle [anti-quasiparticle] with 
momentum ${\bf p}$. Comparing (\ref{curr1}) and (\ref{iik}), we have 
\begin{equation} 
g^{(+)}_B (x; {\bf p}) = N_B^{(+)} (x; {\bf p}) , \;\;\; 
g^{(-)}_B (x; {\bf p}) = - 1 - N_B^{(-)} (x; - {\bf p}) . 
\label{kazu} 
\end{equation} 
It is to be noted that the argument $x$ here is a macroscopic 
spacetime coordinates, which we write $X$ throughout in the 
following. 
\subsubsection{$|p_0|$ prescription} 
Let us write\cite{le} $N_B^{(\pm)} (X, \pm {\bf p}) = N_B^{(\pm)} 
(X; E_p, \pm \hat{\bf p})$, where $\hat{\bf p} \equiv {\bf p} / 
|{\bf p}|$. In the case of ETFT, $N_B^{(\pm)} = N_B^{(\pm)} (E_p)$. 
From (\ref{Deltac-1}), we see that $g^{(\pm)}_B (X, {\bf p})$ 
appears in combination with $\delta (P^2 - m^2) = \delta (p_0^2 - 
E_p^2)$. Then, at first sight, it seems that no difference arises 
between $N_B^{(\pm)} (X; E_p, \pm \hat{\bf p})$ and $N_B^{(\pm)} (X; 
\pm p_0, \pm \hat{\bf p})$. It is well known in 
ETFT that this is not the case, since, in general, $\Delta_c (P)$ is 
to be multiplied by the functions that are singular at $|p_0| = 
E_p$. The correct choice has been known\cite{nie} for some time now 
to be $N_B^{(\pm)} (\pm p_0)$ $(= N_B^{(\pm)} (|p_0|)$ --- the 
\lq\lq $|p_0|$ prescription.'' On the basis of this observation, we 
assume\footnote{It can be shown that, as in ETFT, (\ref{p0f}) is 
consistent with mass-derivative formula\cite{fmuo,nie} at 
least up to the terms with second-order $X^\mu$-derivative.} that 
the $|p_0|$ prescription should be adopted: 
\begin{eqnarray} 
g^{(+)}_B (X; {\bf p}) & = & N_B^{(+)} (X; E_p, \hat{{\bf p}}) \to 
N_B^{(+)} (X; p_0, \hat{\bf p}) \equiv f^{(+)}_B (X; P) , 
\nonumber \\ 
g^{(-)}_B (X; {\bf p}) & = & - 1 - N_B^{(-)} (X; E_p, - \hat{{\bf 
p}}) \to - 1 - N_B^{(-)} (X; - p_0, - \hat{\bf p}) \equiv f^{(-)}_B 
(X; P) , \nonumber \\ 
f_B (X; P) & = & \theta (p_0) \, f^{(+)}_B (X; P) + \theta (- p_0) 
\, f^{(-)}_B (X; P) . 
\label{p0f} 
\end{eqnarray} 

It is to be noted that the translation into the $|p_0|$ prescription 
is {\em formally} achieved by (cf. (\ref{Gc})~-~(\ref{mat})), 
\begin{eqnarray} 
\Delta_c & = & \Delta_R \cdot (1 + 2 f_B) - (1 + 2 f_B) \cdot 
\Delta_A , \nonumber \\ 
\hat{\Delta} & = & \hat{B}_L (f_B) \cdot \hat{\Delta}_{diag} \cdot 
\hat{B}_R (f_B) , 
\label{form} 
\end{eqnarray} 
where $f_B (x, y)$ is defined by 
\[ 
f_B (x, y) = \int \frac{d^{\, 4} P}{(2 \pi)^4} \, e^{- i P \cdot 
(x - y)} f_B (X; P) \;\;\;\;\;\;\; (X = (x + y)/ 2) 
\] 
with $f_B (X; P)$ as in (\ref{p0f}). Now (\ref{const-b}) turns out 
to 
\begin{equation} 
P \cdot \partial_X f_B (X; P) = 0 . 
\label{const} 
\end{equation} 
This is a continuity equation for $f_B$ along the \lq\lq flow line'' 
in a four-dimensional $X^\mu$-space. Equation (\ref{const}) may be 
solved as 
\begin{equation} 
f_B (X; P) = f_B (X_0 = T_{in}, {\bf X} - (X_0 - T_{in}) {\bf p} / 
p_0 ; P) , 
\label{kai} 
\end{equation} 
where $T_{in}$ $(= - \infty)$ is the initial time (cf.~\S2.1). The 
propagator $\hat{\Delta} (X; P)$ with this $f_B (X; P)$ takes the 
form (cf. (\ref{Deltac-1})), 
\begin{equation} 
\Delta_c (X; P) = - 2 \pi i \, \epsilon (p_0) [1 + 2 f_B (X; P)] \, 
\delta (P^2 - m^2) . 
\label{p0-0} 
\end{equation} 
It is to be noted that (\ref{const}) guarantees that all 
higher-order terms in the derivative expansion (cf. (\ref{Deltac})) 
vanish and, in this sense, (\ref{p0-0}) is exact. 
\section{Bare-$N$ scheme} 
Interactions among the fields give rise to reactions taking place in 
a system, which, in turn, causes a nontrivial change in the number 
density of quasiparticles. In this section, within the bare-$N$ 
scheme, we analyze the self-energy-part resummed propagator and 
derive a generalized Boltzmann equation. 
\subsection{Self-energy part} 
We start with noticing that $\sum_{i, \, j = 1}^2 \Sigma_{i j} (x, 
y) = 0$, a proof of which is given in Appendix B. Using this 
property and (\ref{mat}), we obtain 
\begin{eqnarray} 
\hat{\underline{\Sigma}} & \equiv & \hat{B}_R (f_B) \cdot 
\hat{\Sigma} \cdot \hat{B}_L (f_B) = \left( 
\begin{array}{cc}
\Sigma_R & \;\; \Sigma_{off} \\ 
0 & \;\; - \Sigma_A 
\end{array}
\right) , 
\label{koro} \\ 
\Sigma_R & = & \Sigma_{1 1} + \Sigma_{1 2} , \;\;\;\;\;\; 
\Sigma_A = - (\Sigma_{2 2} + \Sigma_{1 2}) , 
\label{aaha} \\ 
\Sigma_{off} & = & \Sigma_{1 2} + \Sigma_{1 2} \cdot f_B - 
\Sigma_{2 1} \cdot f_B + \Sigma_A \cdot f_B - f_B \cdot \Sigma_A 
, 
\label{self2}
\end{eqnarray} 
where $\Sigma_R$ and $\Sigma_A$ are called the retarded and advanced 
self-energy parts, respectively. Going to momentum space, we obtain 
\begin{equation} 
\Sigma_{off} (X; P) \simeq i \left\{ f_B (X; P), \, Re \Sigma_R (X; 
P) \right\} + i \tilde{\Gamma}^{(p)} (X; P) , 
\label{off} 
\end{equation} 
where 
\begin{eqnarray} 
&& \left\{ A (X; P), \, B (X; P) \right\} \equiv \frac{\partial A 
(X; P)}{\partial X_\mu} \frac{\partial B (X; P)}{\partial P^\mu} - 
\frac{\partial A (X; P)}{\partial P^\mu} \frac{\partial B (X; 
P)}{\partial X_\mu} , 
\\ 
&& i \tilde{\Gamma}^{(p)} (X; P) \equiv (1 + f_B (X; P)) 
\Sigma_{1 2} (X; P) - f_B (X; P) \Sigma_{2 1} (X; P) . 
\label{net-sei} 
\end{eqnarray} 
In deriving (\ref{off}), use has been made of the relation 
$\Sigma_A (X; P) = [\Sigma_R (X; P)]^*$, a proof of which is given 
in Appendix B. 
\subsection{Self-energy-part resummed propagator} 
Let us compute the self-energy-part resummed propagator, 
with the aid of the Sch\-win\-ger-Dyson equation, 
\begin{equation} 
\hat{G} = \hat{\Delta} + \hat{\Delta} \cdot \hat{\Sigma} \cdot 
\hat{G} = \hat{G} \cdot \hat{\Sigma} \cdot \hat{\Delta} + 
\hat{\Delta} . 
\label{SD} 
\end{equation} 
Using (\ref{form}) and (\ref{koro}), we obtain 
\begin{eqnarray} 
\hat{G} & = & \hat{B}_L (f_B) \cdot 
\left( 
\begin{array}{cc} 
G_R & \;\; - G_R \cdot \Sigma_{off} \cdot G_A \\ 
0 & \;\; - G_A 
\end{array} 
\right) \cdot \hat{B}_R (f_B) , 
\label{hosi11} 
\\ 
G_{R (A)} & = & \Delta_{R (A)} \cdot \left[ 1 + \Sigma_{R (A)} 
\cdot G_{R (A)} \right] . 
\label{hosi2} 
\end{eqnarray} 
As a matter of course, $G$'s obtained from (\ref{hosi11}) are 
written in the form (\ref{wa})~-~(\ref{rac}) with $G_R$ ($G_A$) for 
$\Delta_R$ ($\Delta_A$) and, for $\Delta_c$, 
\begin{equation} 
G_c = G_R \cdot ( 1 + 2 f_B) - ( 1 + 2 f_B) \cdot G_A - 2 G_R \cdot 
\Sigma_{off} \cdot G_A . 
\label{mom} 
\end{equation} 
Let us compute $G (X; P)$'s ($X = (x + y) / 2$), the Fourier 
transform of $G (x, y)$'s on $x - y$, to the gradient approximation. 
In ETFT, $\Sigma_{off} = 0$. Furthermore, as will be shown below, 
$\Sigma_{off}$ is proportional to $\partial N^{(\pm)} / \partial 
X^\mu$, where $N^{(\pm)}$ are (the contribution to) the physical 
number densities. Then, in the derivative expansion for 
$\Sigma_{off} (x, y)$ (cf. (\ref{hosi})), we keep only zeroth order 
(no-derivative) term. Thus, we obtain 
\begin{eqnarray} 
G_{R (A)} (X; P) & \simeq & \frac{1}{P^2 - m^2 - \Sigma_{R (A)} 
(X; P) \pm i \epsilon (p_0) 0^+} , 
\label{re-re} \\ 
G_c (X; P) & \simeq & G_c^{(0)} (X; P) + G_{c 1}^{(1)} (X; P) 
+ G_{c 2}^{(1)} (X; P) , 
\label{sum3} 
\end{eqnarray} 
where 
\begin{eqnarray} 
G_c^{(0)} (X; P) & = & [1 + 2 f_B (X; P)] \, [G_R (X; P) - G_A 
(X; P)] , 
\label{sum33} \\ 
G_{c 1}^{(1)} (X; P) & = & - \left\{ f_B (X; P), \, Im \Sigma_R 
(X; P) \right\} \left( G_R^2 (X; P) - G_A^2 (X; P) \right) \nonumber 
\\ 
& & - i \tilde{\Gamma}^{(p)} (X; P) \left( G_R^2 (X; P) + 
G_A^2 (X; P) \right) , 
\label{1} \\ 
G_{c 2}^{(1)} (X; P) & = & \Sigma_{off} (X; P) \left( G_R (X; P) - 
G_A (X; P) \right)^2 . 
\label{2} 
\end{eqnarray} 
It is to be noted that (\ref{re-re}) is exact to the gradient 
approximation, i.e., no term including first derivative (with 
respect to $X^\mu$) arises. 

In narrow-width approximation, $Im \Sigma_R = - Im \Sigma_A \to 
- \epsilon(p_0) 0^+$, $G_R G_A$ involved in $G_{c 2}^{(1)}$ develops 
pinch singularity in a complex $p_0$-plane, while $G_c^{(0)}$ and 
$G_{c 1}^{(1)}$ turn out to the well-defined distributions. 

For clarifying the physical meaning of $G_{c 2}^{(1)}$, we compute 
the contribution to the physical-number density through analyzing 
the contribution to the statistical average of the current density 
$\langle j^\mu \rangle$, (\ref{cur-1}) with $G_c$ for $\Delta_c$. 
$G_{c 1}^{(1)}$ as well as $G_c^{(0)} - \Delta_c$ lead to 
perturbative corrections to $\langle j^\mu \rangle$ in (\ref{curr1}) 
due to quantum and medium effects, while $G_{c 2}^{(1)}$ yields the 
large contribution, 
\begin{equation} 
\langle \delta j^\mu (X) \rangle = i \int \frac{d^{\, 4} P}{(2 
\pi)^4} \, P^\mu \, \Sigma_{off} (X; P) \left( G_R (X; P) - G_A (X; 
P) \right)^2 , 
\label{div} 
\end{equation} 
which diverges in the narrow-width approximation. In the next 
subsection, we inspect this large contribution more closely. 
\subsection{Generalized Boltzmann equation} 
The contribution of $G_{c 2}^{(1)}$ to the physical number density, 
$\delta N^{(\pm)}$, is obtained from (\ref{div}) (cf. (\ref{iik})), 
which we carry out in Appendix D1: 
\begin{eqnarray} 
&& \delta N^{(\pm)} (X; \pm \omega_\pm, \pm \hat{\bf p}) = 
\frac{Z_\pm}{2} \frac{- i \Sigma_{off}}{|Im \Sigma_R|} 
\, \rule[-3mm]{.14mm}{8.5mm} 
\raisebox{-2.85mm}{\scriptsize{$\; p_0 = \pm \omega_\pm$}} 
\nonumber \\ 
& & \mbox{\hspace*{20ex}} = \frac{Z_\pm}{2} \tau_\pm \left[ 
\Gamma_\pm^{(p)} + \frac{1}{2 
\omega_\pm} \left\{ f_B, \, Re \Sigma_R \right\} \right]_{p_0 = 
\pm \omega_\pm} , 
\label{iii} 
\\ 
&& \Gamma^{(p)}_\pm =  \frac{- i}{2 \omega_\pm} \left[ (1 + 
N_B^{(\pm)}) \Sigma_{1 2 (2 1)} - N_B^{(\pm)} \Sigma_{2 1 (1 2)} 
\right]_{p_0 = \pm \omega_\pm} . 
\label{net-pro} 
\end{eqnarray} 
Here $\omega_\pm$ $(= \omega_{\pm} (X; \pm {\bf p}))$, 
Eq.~(\ref{Cyo}) in Appendix D, is the energy of the $\pm$ 
(quasi\-par\-ti\-cle/anti-quasiparticle) mode with momentum $\pm 
{\bf p}$, $Z_\pm$ (Eq.~(\ref{Z})) is the wave-function 
renormalization factor and $\tau_\pm \equiv 2 \omega_\pm / |Im 
\Sigma_R (X; \pm \omega_\pm, {\bf p})|$ is the time during which the 
$\pm$ mode damps. $\Gamma_\pm^{(p)}$, Eq.~(\ref{net-pro}), that has 
come from $\tilde{\Gamma}^{(p)}$ $\in - i \Sigma_{off}$ in 
(\ref{net-sei}), is the net production rate of the quasiparticle 
(anti-quasiparticle) of momentum ${\bf p}$ ($- {\bf p}$). In fact, 
$\Gamma^{(p)}_\pm$ is the difference between the production rate and 
the decay rate, so that $\Gamma^{(p)}_\pm$ is the net production 
rate. As mentioned above, in the case of equilibrium system, 
$\Gamma^{(p)}_\pm = 0$ (detailed balance formula). Note that 
(\ref{iii}) diverges in the narrow-width limit, $\tau_\pm \to 
\infty$. 

Recalling the notion of the spacetime cells (cf. \S1), in Appendix 
D2, we derive from (\ref{iii}) a generalized Boltzmann equation for 
$N^{(\pm)} \equiv N_B^{(\pm)} (X; \pm \omega_\pm, \pm \hat{\bf p}) 
+$ \\ 
$\delta N^{(\pm)} (X; \pm \omega_\pm, \pm \hat{\bf p})$ with 
$N_B^{(\pm)}$ the bare number density in \S2: 
\begin{equation} 
\frac{\partial N^{(\pm)}}{\partial X_0} + {\bf v}_\pm \cdot 
\nabla_{\bf X} N^{(\pm)} \simeq \left[ Z_\pm \Gamma_\pm^{(p)} \mp 
\frac{\partial \omega_\pm}{\partial X_\mu} \frac{\partial 
N^{(\pm)}_B (X; \pm p_0, \pm \hat{\bf p})}{\partial P^\mu} 
\right]_{p_0 = \pm \omega_\pm} . 
\label{bare-evo1} 
\end{equation} 
The first term on the right-hand side (RHS) is the collision term, 
while the last term represents the effect due to a change of \lq\lq 
mass.'' Discussion of (\ref{bare-evo1}) will be postponed until \S4. 

The physical-number-density ${\cal N}^{(\pm)}$ may be 
written as 
\begin{equation} 
{\cal N}^{(\pm)} = N^{(\pm)} + \Delta N^{(\pm)} . 
\label{bub} 
\end{equation} 
Here, as mentioned above, $\Delta N^{(\pm)}$, being the contribution 
from $G_c^{(0)} + G_{c 1}^{(1)} - \Delta_c$ $(\in G_c)$, is a \lq\lq 
regular correction.'' 
\subsection{Absence of large contributions} 
In perturbatively computing some quantity, $G_c (X; P)$ appears in 
the form, 
\begin{equation} 
i \int \frac{d^{\, 4} P}{(2 \pi)^4} \, G_c (X; P) {\cal F} (P) , 
\label{kis} 
\end{equation} 
where ${\cal F} (P)$ is the function that is regular at $p_0 = \pm 
\omega_\pm$. $G_c^{(0)} + G_{c 1}^{(1)} \in G_c$, Eqs. 
(\ref{sum3})~-~(\ref{1}), leads to a \lq\lq regular contribution'' 
to (\ref{kis}). Equation (\ref{kis}) with $G_c^{(0)} + G_{c 
2}^{(1)}$ for $G_c$ reads 
\begin{equation} 
i \int \frac{d^{\, 4} P}{(2 \pi)^4} \,\left[ G_c^{(0)} (X; P) + G_{c 
2}^{(1)} (X; P) \right] {\cal F} (P) , 
\label{kiss} 
\end{equation} 
which diverges in the narrow-width limit. Using (\ref{koto}) (in 
Appendix D) in $G_{c 2}^{(1)}$ $(= (G_R - G_A)^2 \Sigma_{off})$, we 
obtain 
\begin{eqnarray} 
\mbox{Eq. (\ref{kiss})} & = & \sum_{\tau = \pm} \int \frac{d^{\, 3} 
p}{(2 \pi)^3} \frac{Z_\pm}{2 \omega_\pm} \left[ \pm [1 + 2 
f_B^{(\pm)} (X; \pm \omega_\pm, {\bf p})] - 
\frac{i \Sigma_{off}^{(\pm)}}{|Im \Sigma_R^{(\pm)}|} \right] {\cal F} 
(\pm \omega_\pm, {\bf p}) \nonumber \\ 
& & + \, ... 
\label{hanamaru} \\ 
& = & \sum_{\tau = \pm} \int \frac{d^{\, 3} p}{(2 \pi)^3} 
\frac{Z_\pm}{2 \omega_\pm} \left[ 1 + 2 \left\{ Z_\pm^{- 1} 
N^{(\pm)} + (1 - Z_\pm^{- 1}) N_B^{(\pm)} \right\} \right] {\cal F} 
(\pm \omega_\pm, {\bf p}) \nonumber \\ 
&& + \, ... \, , 
\label{4.29} 
\end{eqnarray} 
where \lq $...$' stands for \lq\lq regular contributions.'' In 
obtaining (\ref{hanamaru}), we have assumed that the quasiparticles 
here are well defined, i.e., $Im \Sigma^{(\pm)}_R = \mp |Im 
\Sigma^{(\pm)}_R|$. Thus, in terms of $N^{(\pm)}$, there does not 
appear the contribution that diverges in the narrow-width limit. The 
\lq\lq regular contributions'' (\ref{4.29}) are a functional of 
$f_B$, which, if one wants, can be rewritten in terms of $N^{(\pm)}$ 
or ${\cal N}^{(\pm)}$ by using the solution to (\ref{bare-evo1}) and 
(\ref{bub}). Computational procedure, within the bare-$N$ scheme, of 
a generic amplitude is similar to the one within the physical-$N$ 
scheme to be explained in \S4. 
\section{Physical-$N$ scheme} 
\subsection{Preliminary} 
Introducing new functions $f^{(\pm)} (x, y) \equiv \delta (x_0 - 
y_0) \, g^{(\pm)} ({\bf x}, {\bf y}; x_0)$, we {\em redefine} the 
fields, $\hat{\phi}$ and $\hat{\phi}^\dagger$, by (\ref{qu-2}) with 
$\hat{B}_{L (R)} (f^{(\tau)})$ for  $\hat{B}_{L (R)} (f^{(\tau)}_B)$ 
(cf. (\ref{mat})). Then, the fields in this scheme are different 
from the fields in the previous sections. In fact, that $f$ does not 
satisfies (\ref{const-b}) or (\ref{const}) means that (\ref{eq1}) 
does not hold. Then, the interaction-picture field $\phi$ and then 
also $\phi^\dagger$ do not obey the Klein-Gordon equation, 
$(\partial^2 + m^2) \phi \neq 0$, which means that $\hat{\cal L}_0$ 
in (\ref{free-hat}) is not the free hat-Lagrangian. For the purpose 
of finding the correct free hat-Lagrangian $\hat{\cal L}_0'$, we 
proceed as follows: Compute $(\partial^2_x + m^2) \hat{\Delta} (x, 
y)$ with $\hat{\Delta} (x, y)$ as in (\ref{aha}) with (\ref{mat}) 
(with $f^{(\tau)}_B \to f^{(\tau)}$). The result may be written in 
the form, 
\begin{eqnarray*} 
\sum_{\tau = \pm} \hat{D}_\tau \cdot \hat{\Delta}^{(\tau)} 
& \simeq & - 1 , \nonumber \\ 
\hat{D}_\tau (x, z) & \equiv & \hat{\tau}_3 (\partial^2_x + m^2) 
\, \delta^{\, 4} (x - z) + \hat{A} \left[ \ddot{g}^{(\tau)} ({\bf 
x}, {\bf z}; x_0) \right. \nonumber \\ 
&& \left. + 2 \dot{g}^{(\tau)} ({\bf x}, {\bf z}; x_0) 
\frac{\partial}{\partial x_0} - (\nabla_x^2 - \nabla_z^2) g^{(\tau)} 
({\bf x}, {\bf z}; x_0) \right] \delta (x_0 - z_0) , 
\end{eqnarray*} 
where $\dot{g} ({\bf x}, {\bf z}; x_0) \equiv \partial g ({\bf x}, 
{\bf z}; x_0) / \partial x_0$, etc., and $\hat{A}$ is a matrix 
with $A_{i j} = (-)^{i + j}$ $(i, j = 1, 2)$. This means that the 
free hat-Lagrangian 
$\hat{\cal L}_0'$ is 
\begin{equation} 
\hat{\cal L}_0' = - \int d^{\, 4} y \sum_{\tau = \pm} 
\hat{\phi}^{(\tau) \dagger} (x) \hat{D}_\tau (x, y) 
\hat{\phi}^{(\tau)} (y) . 
\label{ff} 
\end{equation} 
Starting with $(\partial^2_y + m^2) \hat{\Delta} (x, y)$, we are led 
to the same $\hat{\cal L}_0'$ as (\ref{ff}) above, as it should be. 
Since ${\cal L}_0' \neq {\cal L}_0$, Eqs. (\ref{ff}) and 
(\ref{free-hat}), the counter term $\hat{\cal L}_c = \hat{\cal L}_0 
- \hat{\cal L}_0'$ appears in the hat-Lagrangian, which yields the 
two-point vertex function, 
\begin{eqnarray*} 
i \hat{{\cal V}} (x, y) & = & i \hat{A} \sum_{\tau = \pm} \left[ 
\ddot{g}^{(\tau)} ({\bf x}, {\bf y}; x_0) + 2 \dot{g}^{(\tau)} ({\bf 
x}, {\bf y}; x_0) \frac{\partial}{\partial x_0} \right. \\ 
&& \left. - (\nabla_x^2 - \nabla_y^2) g^{(\tau)} ({\bf x}, {\bf y}; 
x_0) \right] \delta (x_0 - y_0) \nonumber \\ 
& = & 2 \int \frac{d^{\, 4} P}{(2 \pi)^4} \, e^{- i P \cdot (x - y)} 
\sum_{\tau = \pm} \theta (\tau p_0) \left[ P \cdot \partial_X 
g^{(\tau)} (X; {\bf p}) \right] \hat{A} , 
\end{eqnarray*} 
where $X = (x + y) / 2$. Going to the $|p_0|$ prescription, we 
have (cf. (\ref{p0f})), 
\begin{equation} 
i \hat{{\cal V}} (x, y) \; (\equiv i {\cal V} (x, y) \, \hat{A}) 
\to 2 \int \frac{d^{\, 4} P}{(2 \pi)^4} \, e^{- i P \cdot (x - y)} 
\left[ P \cdot \partial_X f (X; P) \right] \hat{A} . 
\label{ex} 
\end{equation} 
It should be emphasized that this result is exact, not relying on 
the gradient approximation. 

The propagator $\hat{\Delta} (X; P)$ takes the form as in \S2, 
provided that $\Delta_c (X; P)$ is given in (\ref{Deltac}) with 
$g^{(\pm)}_B \to g^{(\pm)}$. Going to the $|p_0|$ prescription, it 
reads 
\begin{eqnarray} 
\Delta_c (X; P) & \simeq & - 2 \pi i \epsilon (p_0) [1 + 2 f (X; P)] 
\, \delta (P^2 - m^2) \nonumber \\ 
&& - 2 i P \cdot \partial_X f (X; P) \left[ \Delta_R^2 (P) + 
\Delta_A^2 (P) \right] \nonumber \\ 
( & \equiv & \Delta_c^{(0)} + \Delta_c^{(1)} ) . 
\label{pd} 
\end{eqnarray} 

Now, $\underline{\Sigma} = \hat{B}_R \cdot \hat{\Sigma} \cdot 
\hat{B}_L$ takes the form (\ref{koro}) with 
\begin{equation} 
\Sigma_{off} = \Sigma_{1 2} + \Sigma_{1 2} \cdot f - \Sigma_{2 1} 
\cdot f + \Sigma_A \cdot f - f \cdot \Sigma_A + {\cal V} . 
\label{aiha} 
\end{equation} 
Fourier transformation leads to 
\begin{equation} 
i \Sigma_{off} (X; P) \simeq \left\{ f (X; P), \, P^2 - m^2 - Re 
\Sigma_R (X; P) \right\} - \tilde{\Gamma}^{(p)} (X; P) . 
\label{r-cond} 
\end{equation} 
The $\hat{\Sigma}$-resummed propagator is written in the form 
(\ref{re-re})~-~(\ref{2}) with $f_B \to f$. 

So far $f$ has not been specified. 
\subsection{\lq\lq Renormalization condition'' and the generalized 
Boltzmann equation} 
We now define $f$, such that the number densities 
\begin{eqnarray} 
N^{(+)} (X; \omega_+, \hat{\bf p}) & = & f (X; p_0 = \omega_+, 
\hat{\bf p}) , \nonumber \\ 
N^{(-)} (X; \omega_-, - \hat{\bf p}) & = & - 1 - f (X; p_0 = - 
\omega_+, \hat{\bf p}) , 
\label{toon} 
\end{eqnarray} 
are as close as the physical-number densities ${\cal N}^{(\pm)}$. As 
seen in \S3, $G_{c 2}^{(1)}$ in (\ref{2}), which is proportional to 
$\Sigma_{off}$, yields a large contribution to the physical-number 
density. Then, as the determining equation for $f$, we 
adopt $\Sigma_{off} (x, y) = 0$ or 
\begin{equation} 
\Sigma_{off} (X; P) = 0 , 
\label{aka} 
\end{equation} 
which we refer to as the \lq\lq renormalization condition'' for the 
number density. It should be emphasized here that this condition is 
by no means unique. $\Sigma_{off}$ is a \lq\lq functional'' of $f$, 
$\Sigma_{off} [f]$. We can adopt $\Sigma_{off} [\tilde{f}] = 0$ $(f 
\neq \tilde{f})$, provided that $\tilde{f} (X; p_0 = \pm \omega_\pm, 
{\bf p}) = f (X; p_0 = \pm \omega_\pm, {\bf p})$. 

In the present scheme with (\ref{aka}), transformed self-energy part 
$\hat{\underline{\Sigma}}$ (cf. Eq.~(\ref{koro})), is diagonal: 
\[ 
\hat{B}_R (f) \cdot \hat{\Sigma} \cdot \hat{B}_L (f) = \left( 
\begin{array}{cc} 
\Sigma_R & 0 \\ 
0 & - \Sigma_A 
\end{array} 
\right) , 
\] 
which is in conformity with the quasiparticle picture. Thus, the 
present scheme is similar in structure to ETFT: Fourier transform of 
$\hat{\Delta} \cdot \hat{\Sigma} \cdot \hat{\Delta}$, etc., are free 
from pinch singularity and then no pinch singularity appears in 
perturbative calculations. Similarly, $G_c (X; P)$ and then also 
$\hat{G} (X; P)$ are free from pinch singularity in the narrow-width 
limit (cf. above after (\ref{2})). 

Let us inspect the physical implication of the \lq\lq 
renormalization condition'' (\ref{aka}). We write $Re \Sigma_R (X; 
P) = M^2 (X) + Re \Sigma_R' (X; P)$, where $M^2 (X)$ is the 
contribution from a tadpole diagram. Note that the tadpole diagram 
does not contribute to $\tilde{\Gamma}^{(p)}$. Equation (\ref{aka}) 
with (\ref{r-cond}) becomes 
\begin{eqnarray} 
2 P \cdot \partial_{X} f (X; P) & \simeq & \tilde{\Gamma}^{(p)} (X; 
P) + \frac{\partial f (X; 
P)}{\partial X^\mu} \frac{\partial Re \Sigma_R' (X; P)}{\partial 
P_\mu} \nonumber \\ 
&& - \frac{\partial f (X; P)}{\partial P_\mu} \frac{\partial 
[M^2 (X) + Re \Sigma_R' (X; P)]}{\partial X^\mu} . 
\label{kik} 
\end{eqnarray} 
Recalling that $N^{(+)} = N^{(+)} (X; p_0 = \omega_+, \hat{\bf 
p})$ [$N^{(-)} = N^{(-)} (X; - p_0 = \omega_-, - {\bf p})$] in 
(\ref{toon}) is the number density of the quasiparticle 
[anti-quasiparticle] with momentum ${\bf p}$ [$- {\bf p}$], it is 
straightforward to show that (\ref{kik}) becomes, on the mass-shell, 
$p_0 = \pm \omega_\pm$, 
\begin{eqnarray} 
& & \frac{\partial N^{(\pm)}}{\partial X_0} + {\bf v}_\pm \cdot 
\nabla_X N^{(\pm)} \pm \frac{\partial \omega_\pm}{\partial 
X_\mu} \frac{\partial N^{(\pm)}}{\partial P^\mu} 
\rule[-3mm]{.14mm}{8.5mm} \raisebox{-2.85mm}{\scriptsize{$\; p_0 = 
\pm \omega_\pm$}} \nonumber \\ 
& & \mbox{\hspace*{5ex}} = \frac{d N^{(\pm)} (X; \omega_\pm 
(X; \pm {\bf p}), \pm \hat{\bf p})}{d X_0} + \frac{d \omega_\pm}{d 
{\bf p}} \cdot \frac{d N^{(\pm)}}{d {\bf X}} - \frac{\partial 
\omega_\pm}{\partial {\bf X}} \cdot \frac{d N^{(\pm)}}{d {\bf 
p}} \nonumber \\ 
& & \mbox{\hspace*{5ex}} \simeq Z_\pm \Gamma_\pm^{(p)} \, 
\rule[-3mm]{.14mm}{8.5mm} \raisebox{-2.85mm}{\scriptsize{$\; p_0 = 
\pm \omega_\pm$}} , 
\label{evo}
\end{eqnarray} 
which is (almost) identical to (\ref{bare-evo1}) in the bare-$N$ 
scheme. $N^{(\pm)} = N^{(\pm)} (X; \omega_\pm, \pm \hat{\bf p})$ 
here is essentially (the main portion of) the relativistic Wigner 
function and (\ref{evo}) is the generalized kinetic or Boltzmann 
equation (cf. Ref.~17)). 
\subsection{Procedure of solving (\ref{kik})} 
Equation (\ref{kik}) determines $f (X; P)$ self-consistently. One 
can also approximately solve (\ref{kik}), order by order, through 
the following iterative procedure. At the RHS of (\ref{kik}), three 
small quantities\footnote{Note that $\partial Re \Sigma_R / \partial 
X$ is a functional of $\partial f / \partial X$.} are involved; the 
coupling constant $\lambda$, $\tilde{\Gamma}^{(p)}$ and $\partial f 
/ \partial X^\mu$. $\tilde{\Gamma}^{(p)}$ measures how the system is 
far from equilibrium. $P^\mu \partial_{X^\mu} f$ has been determined 
by (\ref{kik}) at lower orders (than the order under consideration) 
but the other components of $\partial_{X^\mu} f$ reflect the form of 
initial $f \, \rule[-1.65mm]{.14mm}{4.8mm} 
\raisebox{-1.5mm}{\scriptsize{$\, X_0 = T_{in}$}}$ $( = f_B \, 
\rule[-1.65mm]{.14mm}{4.8mm} \raisebox{-1.5mm}{\scriptsize{$\, X_0 = 
T_{in}$}})$. For simplifying the following presentation, we 
assume\footnote{For the case of $m^2 \lesssim \lambda {\cal T}^2$ 
(with ${\cal T}$ the typical scale(s) of the system) the tadpole 
self-energy-part resummed propagator\cite{pis,eco} should be 
substituted for the bare propagator $\hat{\Delta}$.} that the 
interplay of all these small quantities yields three terms, on the 
RHS of (\ref{kik}), which are of the same order of magnitude. 
[For other cases, (trivial) modification of the following procedure 
is necessary.] Let us write the RHS of (\ref{kik}) ${\cal R} 
[\Delta_c (f)]$, which is a functional of $f$, and expand $f$, 
$\Delta_c$ and ${\cal R}$ (with respect to the small quantity): $f = 
\sum_{i = 0} f^{(i)}$. $\Delta_c = \sum_{i = 0} \Delta_c^{(i)}$, 
${\cal R} = \sum_{i = 1} {\cal R}^{(i)}$. Here $\Delta^{(i)}_c$ are 
as in (\ref{pd}) and $f^{(0)}$ is nothing but $f_B$ in \S2~-~\S3 and 
satisfies (\ref{const}), $P \cdot \partial_X f^{(0)} = 0$, whose 
solution is given in (\ref{kai}). 

\noindent {\em Zeroth order:} Equation (\ref{kik}) reduces to 
(\ref{const}): $P \cdot \partial_X f^{(0)} = 0$.

\noindent {\em First order:} Equation (\ref{kik}) becomes 
\begin{equation} 
2 P \cdot \partial_X f^{(1)} (X; P) = {\cal R}^{(1)} [\Delta_c^{(0)} 
(f^{(0)})] . 
\label{itiji} 
\end{equation} 
Solve (\ref{itiji}) under the condition $f^{(1)} \, 
\rule[-1.65mm]{.14mm}{4.8mm} \raisebox{-1.5mm}{\scriptsize{$\, X_0 = 
T_{in}$}} = 0$. 

\noindent {\em Second order:} Compute $\Sigma_{off} (X; P)$ up to 
second $X^\mu$-derivative terms in the derivative expansion, which 
yields the additional term on the RHS of (\ref{kik}), 
\begin{eqnarray*} 
{\cal R}' [\Delta_c (f)] & = & \frac{i}{8} \left[ \frac{\partial^{\, 
2} (\Sigma_{1 2} - \Sigma_{2 1})}{\partial X^\mu \, \partial X^\nu} 
\frac{\partial^{\, 2} f}{\partial P_\mu \, \partial P_\nu} + 
\frac{\partial^{\, 2} (\Sigma_{1 2} - \Sigma_{2 1})}{\partial P_\mu 
\, \partial P_\nu} \frac{\partial^{\, 2} f}{\partial X^\mu \, 
\partial X^\nu} \right. \\ 
&& \left. - 2 \frac{\partial^{\, 2} (\Sigma_{1 2} - \Sigma_{2 
1})}{\partial X^\mu \, \partial P_\nu} \frac{\partial^{\, 2} 
f}{\partial P_\mu \, \partial X^\nu} \right] . 
\end{eqnarray*} 
Then, we have 
\begin{eqnarray} 
P \cdot \partial_X f^{(2)} & = & {\cal R}^{(2)} [\Delta_c^{(0)} 
(f^{(0)})] + \left\{ {\cal R}^{(1)} [\Delta_c^{(0)} (f^{(0)} + 
f^{(1)}) + \Delta_c^{(1)} (f^{(0)})] \right. \nonumber \\ 
&& \left. - {\cal R}^{(1)} [\Delta_c^{(0)} (f^{(0)})] \right\} + 
{\cal R}' [\Delta_c^{(0)} (f^{(0)})] , 
\label{2ji} 
\end{eqnarray} 
Since $P \cdot \partial_X f^{(0)} = 0$, $\Delta_c^{(1)} [f^{(0)}] = 
0$ (cf. (\ref{pd})). Solve (\ref{2ji}) under the condition $f^{(2)} 
\, \rule[-1.65mm]{.14mm}{4.8mm} \raisebox{-1.5mm}{\scriptsize{$\; 
X_0 = T_{in}$}} = 0$. 

\noindent {\em Higher orders:} Proceed similarly as above. Further 
\lq\lq improvements'' of (\ref{pd}) and (\ref{kik}) are necessary. 

More efficient way is to construct $\hat{\Sigma}$-resummed 
propagator, Eqs. (\ref{re-re})~-~(\ref{1}) ($\Sigma_{off} = 0$), at 
each order and use it for yet higher-order calculations (cf. below). 
\subsection{Procedure of perturbative calculation} 
Perturbative computation, to $N$th order, of an $n$-point amplitude 
${\cal A} (P_1, ... , P_n)$, 
\begin{equation} 
{\cal A} (P_1, ... , P_n) = \int \prod_{i = 1}^n d^{\, 4} x_i \, 
e^{- i \sum_{l} P_l \cdot x_l} \tilde{A} (x_1, ... , x_n) , 
\label{i} 
\end{equation} 
goes as follows. 

1) Draw all relevant skeleton diagrams. Contribution to $\tilde{\cal 
A}$ of each skeleton diagram may be written in the form, 
\begin{equation} 
\int \prod_{i = 1}^j d^{\, 4} z_i \, \tilde{S} (x_1, ... , x_n; 
z_1, ... , z_j) . 
\label{ro} 
\end{equation} 
Here $x_1, ... , x_n$ are the external-vertex (spacetime) points and 
$z_1, ... , z_j$ are the in\-ter\-nal-vertex (spacetime) points, 
which we collectively write $\xi_1, ... , \xi_{n + j}$.  

2) For the propagators involved in each skeleton diagram, use the 
self-energy-part ($\hat{\Sigma}$) resummed propagator $\hat{G}$, 
with $\hat{\Sigma}$ and $f$ computed to appropriate orders. [For 
the skeleton diagram that is already of $N$th order, it is sufficient 
to use the bare propagators and the \lq\lq bare'' $f$'s.] 

3) To analyze $\tilde{S} (\xi_1, ... , \xi_{n + j})$, as (not 
necessarily independent) variables, employ the center-of-mass 
coordinates of $\tilde{\cal A}$, $X$ $(= \sum_{i = 1}^n x_i / n)$, 
and relative coordinates $\xi_{i_k} - \xi_{j_k}$ ($k = 1, ... , m$ 
with $m$ a number of propagators). Here $\xi_{i_k}$ and $\xi_{j_k}$ 
in $\xi_{i_k} - \xi_{j_k}$ are the points that are connected by the 
propagator $\hat{G} (\xi_{i_k}, \xi_{j_k})$ $(\equiv \hat{G}_k)$ , 
\[ 
\hat{G}_k = \hat{G} \left( \frac{\xi_{i_k} + \xi_{j_k}}{2}; 
\xi_{i_k} - \xi_{j_k} \right) = \int \frac{d^{\, 4} Q_k}{(2 \pi)^4} 
\, e^{- i Q_k \cdot (\xi_{i_k} - \xi_{j_k})} \hat{G} \left( 
\frac{\xi_{i_k} + \xi_{j_k}}{2}; Q_k \right) , 
\] 
whose form is known above. [An external as well as an internal point 
appears, in general, in several relative coordinates.] It can easily 
be shown that $(\xi_{i_k} + \xi_{j_k}) / 2$ may be \lq\lq 
decomposed'' as 
\begin{equation} 
\frac{\xi_{i_k} + \xi_{j_k}}{2} = X + \sum_{l \, (\neq k)} 
c_{k l} (\xi_{i_l} - \xi_{j_l}) , 
\label{A} 
\end{equation} 
where $c$'s are numerical coefficients. The decomposition (\ref{A}) 
is not unique, which is not the matter for our purpose. Carry out 
the derivative expansion: 
\begin{equation} 
\hat{G}_k = \int \frac{d^{\, 4} Q_k}{(2 \pi)^4} \, e^{- i Q_k 
\cdot (\xi_{i_k} - \xi_{j_k})} \left[ 1 + \sum_{l \, (\neq k)} 
c_{k l} (\xi_{i_l} - \xi_{j_l} ) \cdot \partial_X + ... \right] 
\hat{G} (X; Q_k)) . 
\label{C} 
\end{equation} 
Pick up the term with $(\xi_{i_l} - \xi_{j_l})$. Multiplication of 
$\hat{G}_l$ $(\in \tilde{\cal S})$ yields 
\begin{eqnarray*} 
&& c_{k l} (\xi_{i_l} - \xi_{j_l}) \cdot \partial_X \hat{G} (X; Q_k) 
\, \hat{G}_l \nonumber \\ 
&& \;\;\; = c_{k l} \partial_{X_\mu} \hat{G} (X; Q_k) \int 
\frac{d^{\, 4} Q_l}{(2 \pi)^4} \, e^{- i Q_l \cdot (\xi_{i_l} - 
\xi_{j_l})} \left( \frac{1}{i} \frac{\partial}{\partial Q_l^{\mu}} 
\right) \hat{G} \left( \frac{\xi_{i_l} + \xi_{j_l}}{2}; Q_l \right) 
. 
\end{eqnarray*} 
The higher $X^\mu$-derivative terms \lq $...$' in (\ref{C}) may be 
dealt with similarly. Deal with all the propagators in the above 
manner. For $f (X; Q_l)$'s, substitute the solution (of appropriate 
order) to (\ref{aka}). 

4) Carry out the integrations over $z$'s in (\ref{ro}), which yields 
$j$ momentum-conser\-va\-tion $\delta$-functions at the internal 
points, $z$'s. 

5) Carry out the integrations over $j$ $Q$'s. Add the contributions 
of all the diagrams to obtain $\tilde{A} (x_1, ... , x_n)$, which, 
as is obvious from the above procedure, is of the form $\tilde{A} 
(X; x_1 - x_2, ... , x_{n - 1} - x_n)$.   

6) Substitute $\tilde{\cal A}$ thus obtained into (\ref{i}) to 
obtain 
\begin{eqnarray} 
{\cal A} (P_1, ... , P_n) & = & \int d^{\, 4} X \, A (X; P_1, ... , 
P_n) , 
\label{ti} \\ 
A (X; P_1, ... , P_n) & = & \int \prod_{i = 1}^{n - 1} d^{\, 4} 
(x_{i + 1} - x_i) \, e^{- i \sum_{i = 1}^{n - 1} P_i' \cdot (x_{i + 
1} - x_i)} \nonumber \\ 
&& \times \tilde{A} (X; x_1 - x_2, ... , x_{n - 1} - x_n) , 
\end{eqnarray} 
where $P_i'$ is the linear combination of $P_1, ... , P_n$. Carry 
out the integration over $x_{i + 1} - x_i$ $(i = 1, ... , n - 1)$ to 
obtain $A (X; P_1, ... , P_n)$. Note that the momentum conservation 
$\sum_{i = 1}^n P_i = 0$ holds (cf., however, the footnote in 
\S2.2.2). Note also that $A$ depends weakly on $X$ through $f (X; 
Q_k)$'s. Were it not for this $X$-dependence, integration over $X$ 
in (\ref{ti}) would yields $(2 \pi)^4 \delta^{\, 4} (\sum_i P_i)$. 
\section{Comparison with related work} 
As has been mentioned in \S1, the physical-$N$ scheme is employed 
in the literature. 
\subsection{CTP formalism} 
The derivation of the generalized Boltzmann equation (GBE) in 
Refs.~2), 7), 18) and 19)\footnote{In Ref.~19), nonrelativistic 
many-body theory is dealt with.} starts with the Schwinger-Dyson 
equation, Eq.~(\ref{SD}), 
\begin{equation} 
\left( \hat{\Delta}^{- 1} - \hat{\Sigma} \right) \hat{G} = \hat{G} 
\left( \hat{\Delta}^{- 1} - \hat{\Sigma} \right) = 1 . 
\label{SD2} 
\end{equation} 
Here we stress that this equation is nothing more than 
the equation 
that serves as resumming the self-energy part to makeup the resummed 
propagator, (\ref{hosi2}) and (\ref{mom}). Then in order to derive 
the GBE, an additional input or condition is necessary. Our 
condition is the \lq\lq renormalization condition'' $\Sigma_{off}$ 
in (\ref{aka}). 

We first see the derivation in Refs.~7), 18) and 19), in which the 
lowest nontrivial order is dealt with. 
The condition adopted there is essentially 
\begin{equation} 
G_c (X; P) = [1 + 2 f' (X; P)] \, [G_R (X; P) - G_A (X; P) ] . 
\label{tyu} 
\end{equation} 
[To distinguish from ours, we write $f'$ for $f$ here.] No counter 
Lagrangian is introduced, so that the solution for $G_c$ to 
(\ref{SD2}) is given by (\ref{mom}) with $f_B \to f'$ (cf. the 
argument in \S4): 
\begin{equation} 
G_c = G_R \cdot ( 1 + 2 f') - ( 1 + 2 f') \cdot G_A - 2 G_R \cdot 
\Sigma_{off} \cdot G_A , 
\label{mom1} 
\end{equation} 
where $\Sigma_{off}$ is as in (\ref{self2}) with $f_B \to f'$. 
Note that the condition (\ref{tyu}) is the leading term of the 
(Fourier transform of the) exact form (\ref{mom1}) 
with $\Sigma_{off = 0}$. 

Fourier transforming (\ref{mom1}) and equating with (\ref{tyu}), one 
obtains 
\begin{equation} 
\left\{ f' (X; P) , \, G_R (X; P) + G_A (X; P) \right\} \simeq - 2 i 
G_R (X; P) \Sigma_{off} (X; P) G_A (X; P) \, . 
\label{yare} 
\end{equation} 
To the lowest nontrivial order under consideration, the left-hand 
side (LHS) of (\ref{yare}) is approximated as 
\begin{eqnarray} 
\mbox{LHS of (\ref{yare})} & \sim & - 2 i P \cdot \partial_X f' (X; 
P) \{ G_R^2 (X; P) + G_A^2 (X; P) \} \nonumber \\ 
& = &  4 i P \cdot \partial_X f' (X; P) G_R (X; P) G_A (X; P) 
\;\;\;\;\;\;\; 
(p_0 = \pm \omega_\pm) \, . 
\label{yareyare} 
\end{eqnarray} 
Using (\ref{yareyare}) in (\ref{yare}) on the mass-shell, $p_0 = \pm 
\omega_\pm$, one gets the Boltzmann equation. 

In Ref.~2) is given an all-order derivation of the GBE without 
introducing the counter Lagrangian. Although the correct GBE comes 
out, the derivation is not self consistent. In order to see this, we 
first note that, upon using (\ref{aaha}), (\ref{self2}) and 
(\ref{hosi2}), the solution (\ref{mom1}) may be rewritten as 
\begin{eqnarray} 
G_R^{- 1} G_c G_A^{- 1} & = & \tilde{G}_1 + \tilde{G}_2 , 
\label{GGG} 
\\ 
\tilde{G}_1 & = & \Sigma_{1 1} + \Sigma_{2 2} 
\label{G1} \\ 
\tilde{G}_2 & = & 2 \left( f' \cdot \Delta_A^{- 1} - \Delta_R^{- 1} 
\cdot f' \right) . 
\label{G2} 
\end{eqnarray} 
At an intermediate step of solving Schwinger-Dyson equation, there 
appears the term proportional to $\Delta_R^{- 1} \cdot \Delta_c 
\cdot \Delta_R^{- 1}$, which turns out to $\tilde{G}_2$. In Ref.~2), 
$\tilde{G}_2$ is missing, and thus, when interaction is switched 
off, $G_c$ does not reduce to the free form (\ref{form}). [It is to 
be noted that, in ETFT, Fourier transform of $\Delta_R^{- 1} \cdot 
\Delta_c \cdot \Delta_R^{- 1}$ is proportional to $(P^2 - m^2)^2 
\delta (p^2 - m^2)$ and then vanishes.] As the additional condition, 
in place of (\ref{tyu}), $G_c = G_R \cdot (1 + f') - (1 + f') \cdot 
G_A$ is adopted. This again contradicts with the exact solution 
(\ref{mom1}) unless $\Sigma_{off} = 0$. By equating two relations 
above, the correct GBE comes about. 

As has been shown in \S4, for constructing a consistent physical-$N$ 
scheme, the counter Lagrangian ${\cal L}_c$ should be introduced. 
As mentioned above, however, ${\cal L}_c$ is not introduced in 
Refs.~2), 7), 18) and 19), so that no self-consistent 
perturbative calculational procedure is presented. 
\subsection{Thermo field dynamics (TFD)} 
In contrast to the CTP formalism, the TFD counterparts of $\phi_1$ 
and $\phi_2$ are independent fields and mutually commutable. 

We first consider the equilibrium case. In spite of the fact 
mentioned above, as far as the perturbation theory is concerned, TFD 
is equivalent to CTP formalism. However, we should mention one point 
--- the $|p_0|$ prescription (\S2.2.4). In the CTF formalism, the 
$|p_0|$ prescription is deduced\cite{nie} from first principles. 
The very existence of the vertical segments\cite{lan,eco,nie} of 
the time-path in a complex time plane plays an important role. In 
TFD, however, there is no counterpart of the vertical segments, and 
then, on the basis of the consistency argument (with the spectral 
representation and KMS condition), the $|p_0|$ prescription is 
assumed.\cite{umt,fmuo} 

The nonequilibrium TFD\cite{ume,ume1} (NETFD) is formulated on the 
basis of (nonequilibrium generalization of) the so-called 
thermal-state condition. As a consequence of this, initial 
correlations (cf.~(\ref{shoki}) and Appendix A) are absent. Thus, 
when compared to the (CTP) formalism developed in this paper, NETFD 
may be applied to more restricted class of nonequilibrium systems, 
i.e., the systems which are described by suitably defined 
quasiparticles, in terms of which the generalized thermal-state 
condition holds. On the other hand, NETFD is a \lq\lq 
single-(space)time'' canonical formalism, without distinction 
between the microscopic- and 
macroscopic-(space)times. In this sense, NETFD has much wider 
applicability than the CTP formalism presented in this paper, the 
latter of which applies only for \lq\lq out-of-equilibrium 
systems.'' Incidentally, in NETFD, there is one parameter 
($s$-parameter), which has no counter part in the CTP formalism. 

Our condition for determining the \lq\lq physical-number density,'' 
$N^{(\pm)}$, is that, with $N^{(\pm)}$, 
pinch singularities (in the 
narrow-width limit) disappear. On the other hand, NETFD, in which 
the (space)\-time representation is employed, imposes an \lq\lq 
on-shell renormalization condition,'' which results in the GBE. 
Since \lq\lq the pinch singularity'' is a notion in momentum space, 
it seems not to be immediately obvious how to translate this 
condition to the (space)\-time representation as adopted in NETFD. 
Nevertheless, closer inspection of the structure of both formalisms 
tells us that our condition is in accord with the \lq\lq on-shell 
renormalization condition'' in NETFD. Incidentally, how to reconcile 
the NETFD with the $|p_0|$ prescription, a notion in momentum space, 
remains to be an open question. 
\section{Summary and discussion} 
In this paper, we have dealt with perturbative framework of 
out-of-equilibrium relativistic complex-scalar-field systems. 
We have assumed the existence of two different spacetime scales, the 
microscopic and macroscopic. The first small scale characterizes the 
microscopic correlations and the second large scale is inherent in 
the relaxation phenomena. 

We have proposed mutually equivalent two computational schemes, the 
bare-$N$ and physical-$N$ schemes. Both of them lead to the 
generalized relativistic kinetic or Boltzmann equation. Then, we 
have presented the procedure of perturbatively computing a generic 
amplitude. 

It is worth pointing out a similarity between two computational 
schemes presented in this paper and those in UV-renormalization in 
vacuum theory. Taking mass renormalization as an example, let us see 
this. 

A \lq\lq bare scheme'' in vacuum theory takes ${\cal L}_0 = - 
\phi^\dagger (\partial^2 + m_B^2) \phi$, with $m_B$ the bare mass, 
as the free Lagrangian. Then, the propagator reads, $\Delta (P) = 1 
/ (P^2 - m_B^2 + i0^+)$. Perturbative computation of the 
renormalized mass $m$ yields $m = M (m_B)$, which may not 
necessarily be the physical mass ${\cal M}$. Perturbative 
computation of some quantity, e.g., the physical mass ${\cal M}$, 
yields the result as a function of $m_B$, which may be rewritten, 
using $m = M (m_B)$, in terms of $m$. The correspondence of the 
\lq\lq bare scheme'' here to the bare-$N$ scheme in \S2~-~\S3 is as 
follows: ${\cal L}_0 \to \hat{\cal L}_0$ in (\ref{free-hat}), 
$\Delta (P) \to \hat{\Delta} (X; P)$ (Eqs. 
(\ref{reta})~-~(\ref{rac}) with (\ref{Delta-ra}) and (\ref{p0-0})), 
$m_B \to N_B^{(\pm)}$, $m \to N^{(\pm)}$, $m = M (m_B) \to 
Eq.~(\ref{iii})$, ${\cal M} \to {\cal N}^{(\pm)}$. 

A \lq\lq physical scheme'' in vacuum theory takes ${\cal L}_0' = - 
\phi^\dagger (\partial^2 + m^2) \phi$, with $m$ the renormalized 
mass, as the free Lagrangian and then there emerges the 
counter-Lagrangian, ${\cal L}_c = \phi^\dagger (m^2 - m_B^2) \phi$. 
The propagator is $\Delta (P) = 1 / (P^2 - m^2 + i 0^+)$. $m^2 - 
m_B^2$ is determined so that the perturbatively computed mass, $M 
(m, m^2 - m_B^2)$, is equal to $m$, $m = M (m, m^2 - m_B^2)$. The 
correspondence of the \lq\lq physical scheme'' here to the 
physical-$N$ scheme in \S4 is as follows: ${\cal L}_0' \to 
\hat{\cal L}_0'$ in (\ref{ff}), $\Delta (P) \to \hat{\Delta} (X; P)$ 
(Eqs. (\ref{reta})~-~(\ref{rac}) with (\ref{Delta-ra}) and 
(\ref{pd})), ${\cal L}_c \to \hat{\cal L}_c = \hat{\cal L}_0 - 
\hat{\cal L}_0'$, $m = M (m, m^2 - m_B^2) \to \Sigma_{off} = 0$ in 
(\ref{aka}), $m \to N^{(\pm)}$, ${\cal M} \to {\cal N}^{(\pm)}$. 

To summarize, the bare-$N$ scheme is constructed in terms of the 
original bare-number density $f_B (X; P)$ [respect. the bare mass 
$m_B$]. On the other hand, the physical-$N$ scheme is constructed in 
terms of the \lq\lq renormalized''-number density $f (X; P)$ 
[respect. the renormalized mass $m$]. Both perturbative schemes are 
equivalent. The first scheme starts with the \lq\lq bare quantity'' 
and the \lq\lq renormalization'' is done at the end, while in the 
second scheme, the \lq\lq renormalization'' is done at the beginning 
by introducing the counter hat-Lagrangian $\hat{\cal L}_c$ [respect. 
${\cal L}_c$]. It is worth recalling that the renormalized mass $m$ 
is defined so as to absorb UV divergences. However, there is 
arbitrariness in the definition of the \lq\lq finite part'' of $m$, 
which is determined by imposing some condition. Under this 
condition, $m$ is determined order by order in perturbation series. 
This is also the case in the present physical-$N$ scheme. $f$ (or 
$N^{(\pm)}$) is defined so as to absorb large contributions, which 
diverge in narrow-width limit. As pointed out above after 
(\ref{aka}), there is arbitrariness in fixing the \lq\lq finite 
part'' of $f$. In \S4, we have imposed the condition for determining 
$f$, under which $f (X; P)$ turns out to be determined order by 
order in perturbation 
series. 
\section*{Acknowledgments}
This work was supported in part by the Grant-in-Aide for 
Sci\-en\-tif\-ic Re\-search ((A)(1) (No.~08304024)) of the Ministry 
of Education, Science and Culture of Japan. 
\setcounter{equation}{0}
\setcounter{section}{1}
\section*{Appendix A: Initial correlations} 
\def\theequation{\mbox{\Alph{section}.\arabic{equation}}}
Here we discuss to what extent, the initial correlations ${\cal 
C}_{2 n}$, Eq.~(\ref{shoki}), may be ignored. 

We first note that, in the two-component representation (cf. \S2.1), 
${\cal C}_n$ has $2^n$ components, all of which are 
identical.\cite{chou} Then, from (\ref{reta}) and (\ref{adva}), 
we see that ${\cal C}_2$, which constitutes the medium part of the 
propagator $\hat{\Delta}$, does not appear in the retarded and 
advanced Green functions. This means that ${\cal C}_2$ does not 
appear in the two-point correlation in the linear response theory. 
With the aid of the standard formulae (cf. Sections 2 and 5 of 
Ref.~2)), one can also show that ${\cal C}_{2 n}$ does not 
appear in multi-point correlations in the nonlinear as well as the 
linear response theory. 

We decompose $\phi (x)$ as in vacuum theory: 
\begin{equation} 
\phi (x) = \int \frac{d^{\, 3} q}{\sqrt{ 2 E_q (2 \pi)^3}} \left[ 
a ({\bf q}) e^{- i Q \cdot x} + b^\dagger ({\bf q}) e^{i Q \cdot x} 
\right] , 
\label{11} 
\end{equation} 
where $q_0 = E_q =$ $\sqrt{q^2 + m^2}$. $a ({\bf q})$ 
[$b^\dagger 
({\bf q})$] in (\ref{11}) is the annihilation [creation] operator of 
a particle [antiparticle] in vacuum theory: $[a ({\bf q}) , 
a^\dagger ({\bf q}') ] = [b ({\bf q}) , b^\dagger ({\bf q}') ] = 
\delta ({\bf q} - {\bf q}')$. 

We restrict our concern to ${\cal C}_4$. The argument below may be 
generalized to the case of general ${\cal C}_{2 n}$ $(n \geq 3)$. 
We are concerned only with the particle part of $\phi$, the part 
with $a ({\bf q})$ in (\ref{11}), and of $\phi^\dagger$, the 
hermitian conjugate of $\phi$. Other parts may be dealt with 
similarly and the same conclusion will be obtained. 

Substituting (\ref{11}) and its hermitian conjugate to ${\cal C}_4$ 
in (\ref{shoki}) and carrying out the Fourier transformation, we 
obtain, up to an numerical factor, 
\begin{eqnarray} 
&& \tilde{\cal C}_4 \propto \frac{1}{\prod_{j = 1}^2 \sqrt{E_{p_j} 
E_{q_j}}} \prod_{j = 1}^2 \left[ \delta (p_j^0 - E_{p_j}) \, \delta 
(q_j^0 - E_{q_j}) \right] \langle {\bf q}_1, {\bf q}_2; {\bf p}_1, 
{\bf p}_2 \rangle_c 
\label{kob} \\ 
&& \langle {\bf q}_1, {\bf q}_2 ; {\bf p}_1, {\bf p}_2 \rangle_c 
\equiv \langle {\bf q}_1, {\bf q}_2; {\bf p}_1, {\bf p}_2 \rangle - 
\langle {\bf q}_1; {\bf p}_1 \rangle \langle {\bf q}_2; {\bf p}_2 
\rangle - \langle {\bf q}_1; {\bf p}_2 \rangle 
\langle {\bf q}_2; {\bf p}_1 \rangle , \nonumber 
\end{eqnarray} 
where $\langle {\bf q}_1, {\bf q}_2 ; {\bf p}_1, {\bf p}_2 \rangle 
\equiv \langle a^\dagger ({\bf q}_1) a^\dagger ({\bf q}_2) a ({\bf 
p}_1) a ({\bf p}_2) \rangle$ and $\langle {\bf q}; {\bf p} \rangle 
\equiv \langle a^\dagger ({\bf q}) a ({\bf p}) \rangle$, etc. The 
setup in \S1, ${\cal C}_4$ does not change appreciably in a single 
spacetime cell, leads to approximate momentum-conservation: 
\begin{equation} 
\tilde{C}_4 \simeq 0 \;\;\; \mbox{for} \;\, \sum_{j = 1}^2 
(P_j^\mu - Q_j^\mu) \gtrsim 1 / L^\mu . 
\label{m-c} 
\end{equation} 
Thus, we may write 
\begin{equation} 
\tilde{C}_4 = \Delta^{\, 4} \left( \sum_{j = 1}^2 (P_j - Q_j) 
\right) \prod_{j = 1}^2 \left[ \delta_+ (P_j^2 - m^2) \, \delta_+ 
(Q_j^2 - m^2) \right] \, {\cal A} , 
\label{n-c} 
\end{equation} 
where $\Delta (P^\mu)$ is the function whose width [height] is of $O 
(1 / L^\mu)$ [$O (L^\mu)$] and satisfies $\int \Delta (P^\mu) d 
P^\mu = 1$, and $\delta_+ (P^2 - m^2) \equiv \theta (p_0) \delta 
(P^2 - m^2)$. 

A transition probability or a rate of a microscopic reaction is 
related to an on-shell amplitude. (In the case of equilibrium 
thermal field theories, this relation is settled in 
Ref.~20).) Let us now analyze the on-shell ($P_j^2 = m^2$, etc.) 
amplitudes. As can be seen from (\ref{kob}), ${\cal C}_4$ {\em per 
se} does not contribute to the on-shell amplitudes. Then, ${\cal 
C}_4$ or $\tilde{\cal C}_4$ appears as a part(s) of an on-shell 
amplitude. Let us see the structure of a connected amplitude $S$, 
which includes a single ${\cal C}_4$. We may write, with obvious 
notation, 
\begin{eqnarray} 
S & = & \int \prod_{j = 1}^2 \left[ d^{\, 4} P_j \, d^{\, 4} Q_j 
\right] \tilde{C}_4 \, {\cal F} (Q_1, Q_2; P_1, P_2) \nonumber \\ 
& \propto & \int \frac{d^{\, 3} p_1 d^{\, 3} p_2 d^{\, 3} 
q_1}{E_{p_1} E_{p_2} E_{q_1} E_{|{\bf p}_1 + {\bf p}_2 - 
{\bf q}_1|}} \Delta (E_{p_1} + E_{p_2} - E_{q_1} - E_{|{\bf p}_1 
+ {\bf p}_2 - {\bf q}_1|}) \nonumber \\ 
&& \times {\cal A} \, {\cal F} (P_1, P_2; Q_1, Q_2 = P_1 + P_2 - 
Q_1) . 
\label{kann} 
\end{eqnarray} 

$\tilde{C}_4$ is to be compared with the four-point Green functions 
$\tilde{G}_4$'s. For the sake of comparison, we take the $n$th-order 
contribution, $\tilde{G}_4^{(n)}$, whose four legs are bare 
propagator with on-shell $\delta$-function: 
\[ 
\tilde{G}_4^{(n)} \propto \lambda^{n + 1} \delta^4 \left( 
\sum_{j = 1}^2 (P_j - Q_j) \right) \prod_{j = 1}^2 \left[ \delta 
(P_j^2 - m^2) \, \delta (Q_j^2 - m^2) \right] {\cal B} ,  
\] 
The counterpart of $S$, Eq.~(\ref{kann}), is 
\begin{eqnarray} 
\tilde{S}^{(n)} & = & \lambda^{n + 1} \int \frac{d^{\, 3} p_1 d^{\, 
3} p_2 d^{\, 3} q_1}{E_{p_1} E_{p_2} E_{q_1} E_{|{\bf p}_1 + {\bf 
p}_2 - {\bf q}_1|}} \delta (E_{p_1} + E_{p_2} - E_{q_1} - 
E_{|{\bf p}_1 + {\bf p}_2 - {\bf q}_1|}) \nonumber \\ 
&& \times {\cal B} \, {\cal F} 
(P_1, P_2; Q_1, Q_2 = P_1 + P_2 - Q_1) . 
\label{haha} 
\end{eqnarray} 
From (\ref{kann}) and (\ref{haha}), we obtain, with obvious 
notation, 
\begin{equation} 
\frac{S}{\tilde{S}^{(n)}} = \frac{\langle {\cal A} \, {\cal F} 
\rangle}{\lambda^{n + 1} \langle {\cal B} \, {\cal F} \rangle} . 
\label{haha1} 
\end{equation} 
Noticing that ${\cal A}$ and ${\cal B}$ are dimensionless, we assume 
that ${\cal A}$ and ${\cal B}$ are of $O (1)$. Equation (\ref{haha1}) 
tells us that, unless $\langle {\cal A} {\cal F} \rangle / 
\langle {\cal B} {\cal F} \rangle << 1 / \lambda^{n + 1}$, ${\cal 
C}_4$ may not be ignored.  

Equation (\ref{A1}) in Appendix C, together with the setup in \S1, 
tells us that 
\begin{equation} 
\langle a^\dagger ({\bf q}) a ({\bf p}) \rangle \simeq 0 
\;\;\;\;\; \mbox{for} \;\, |{\bf q} - {\bf p}| \gtrsim 1 / |{\bf L}| 
. 
\label{oku} 
\end{equation} 
Motivated with this relation, we consider a case, where $\langle 
{\bf q}_1, {\bf q}_2; {\bf p}_1, {\bf p}_2 \rangle_c$ has the 
property of factorizability or short-range correlation in momentum 
space: 
\begin{equation} 
\langle {\bf q}_1, {\bf q}_2; {\bf p}_1, {\bf p}_2 \rangle_c 
\simeq 0 \;\;\;\;\; \mbox{for} \;\, |{\bf p}_1 - {\bf q}_1| \gtrsim 
1 / |{\bf L}| \; \mbox{and} \; |{\bf p}_1 - {\bf q}_2| \gtrsim 1 / 
|{\bf L}| . 
\label{uee} 
\end{equation} 
In this case, using (\ref{uee}) in (\ref{kob}), we see that 
(\ref{kann}) becomes, with obvious notation, 
\begin{equation} 
S \propto \frac{1}{V} \int \frac{d^{\, 3} p_1 d^{\, 3} 
p_2}{E_{p_1}^2 E_{p_2}^2} \Delta (E_{p_1} + E_{p_2} - E_{q_1} - 
E_{|{\bf p}_1 + {\bf p}_2 - {\bf q}_1|}) \, {\cal A} \, {\cal F} 
\, \rule[-1.9mm]{.14mm}{5.4mm} \raisebox{-1.5mm}{\scriptsize{$\; 
{\bf p}_1 \simeq {\bf q}_1 \oplus {\bf p}_2 \simeq {\bf q}_1$}} , 
\label{oww} 
\end{equation} 
where $V = L^1 L^2 L^3$. On the basis of the observation made at the 
footnote in \S2.2.2, we assume that $\tilde{S}^{(n)}$ in 
(\ref{haha}) is insensitive to the region $|{\bf q}_1| \lesssim 1 / 
|{\bf L}|$. Then, in (\ref{haha}), the integration over ${\bf q}_1$ 
yields an $O ({\cal T}^3)$ quantity, with ${\cal T}$ the typical 
scale(s) of the system, which characterizes the microscopic reaction 
and is much larger than $1 / |{\bf L}|$. Since we have assumed that 
${\cal A}$ and ${\cal B}$ are of $O (1)$, we obtain 
\[ 
\frac{S}{\tilde{S}^{(n)}} = O \left( \frac{1}{\lambda^{n + 1} {\cal 
T}^3 \, V} \right) . 
\] 
Thus, $S / \tilde{S}^{(n)} << 1$ and the initial correlation 
${\cal C}_4$ may be ignored. There are, however, several cases, in 
which this observation does not apply: 
\begin{itemize} 
\item Dimensionless quantity ${\cal A}$ in (\ref{oww}) is large. 
\item ${\cal F}$ in (\ref{oww}) is extremely large at ${\bf q}_1 
\simeq {\bf p}_1$ and/or ${\bf q}_1 \simeq {\bf p}_2$. 
\item In massless theory $(m = 0)$, some quantities are sensitive to 
the infrared region (cf. the footnote in \S2.2.2). If $S$ is 
(\ref{kann}) is such a quantity, more elaborate analysis is 
necessary to see if ${\cal C}_4$ may be ignored or not. 
\end{itemize} 
\setcounter{equation}{0}
\setcounter{section}{2}
\def\theequation{\mbox{\Alph{section}.\arabic{equation}}}
\section*{Appendix B: Some properties of the propagator and the 
self-energy part} 
Self-energy-part ($\hat{\Sigma}$) inserted propagator is obtained 
by perturbatively computing the formula for the full propagator: 
\begin{equation} 
\hat{\cal G} (x, y) \equiv - i \langle T_C \left( \hat{\phi} (x) 
\hat{\phi}^\dagger (y) e^{i \int d^{\, 4} z \, \hat{\cal L}_{int} 
(z)} \right) \rangle , 
\label{huu} 
\end{equation} 
where $\hat{\cal L}_{int}$ $( = \hat{\cal L} - \hat{\cal L}_0)$ is 
the interaction hat-Lagrangian. From this, we obtain ${\cal G}_{1 1} 
+ {\cal G}_{2 2} = {\cal G}_{1 2} + {\cal G}_{2 1}$, of which 
(\ref{wa}) is a special case. Applying this to the 
single-$\hat{\Sigma}$ inserted propagator ($\hat{\cal G} \to 
\hat{\Delta} \cdot \hat{\Sigma} \cdot \hat{\Delta}$) and using 
(\ref{wa}), we obtain $\sum_{i, \, j = 1}^2 \Sigma_{i j} (x, y) = 
0$. 

Taking complex-conjugate of (\ref{huu}), we see that 
\[ 
i {\cal G}_{1 1} (x, y) = \left[ i {\cal G}_{2 2} (y, x) 
\right]^* , \;\;\;\;\;\; i {\cal G}_{1 2 (2 1)} (x, y) = 
\left[ i {\cal G}_{1 2 (2 1)} (y, x) \right]^* . 
\] 
Using this for $\hat{\cal G} \to \hat{\Delta}$ and $\hat{\cal G} \to 
\hat{\Delta} \cdot \hat{\Sigma} \cdot \hat{\Delta}$, we obtain 
\[ 
- i (\Sigma (x, y))_{1 1} = \left[ - i (\Sigma (y, x))_{2 2} 
\right]^* , \;\;\;\; - i (\Sigma (x, y) )_{1 2 (2 1)} = \left[ - 
i (\Sigma (y, x))_{1 2 (2 1)} \right]^* . 
\] 
Fourier transformation on $x - y$ yields 
\begin{eqnarray*} 
& & (\Sigma (X; P) )_{1 1} = - (\Sigma (X; P) )_{2 2}^* , 
\nonumber \\ 
& & (\Sigma (X; P) )_{1 2} \, ,\;\;\; (\Sigma (X; P) )_{2 1} \; : 
\mbox{pure imaginary.} 
\end{eqnarray*} 
Using this in (\ref{aaha}), we obtain $\Sigma_A (X; P) = [\Sigma_R 
(X; P)]^*$. 

In the case of physical-$N$ scheme, $\hat{\cal L}_c$ in \S4 
contributes to $\hat{\Sigma}$, $- \hat{\cal V} (x, y)$ $\in$ 
$\hat{\Sigma} (x, y)$ (cf. (\ref{ex})). This \lq\lq additional'' 
contribution does not invalidate the above properties. 
\setcounter{equation}{0}
\setcounter{section}{3}
\section*{Appendix C: Representation of $\Delta_c$ in terms of a 
vac\-u\-um-theory kit} 
\def\theequation{\mbox{\Alph{section}.\arabic{equation}}}
Here obtain the expression for $\Delta_c$ and the number density 
$N_B^{(\pm)}$ in terms of the quantities in vacuum theory. 

Straightforward but a bit lengthy calculation using (\ref{11}) and 
its hermitian conjugate yields 
\begin{eqnarray} 
i \Delta_c (X; P) & = & \int d^{\, 4} (x - y) e^{i P \cdot (x - y)} 
\langle \phi(x) \phi^\dagger (y) + \phi^\dagger (y) \phi(x) \rangle 
\nonumber \\ 
& = & i \Delta_c^{(0)} (P) + i \Delta_c^{(1)} 
(X; P) , \\ 
i \Delta_c^{(0)} (P) & = & \frac{\pi}{E_p} \left[ \delta (p_0 - 
E_p) + \delta (p_0 + E_p) \right] \nonumber \\ 
& = & 2 \pi \delta (P^2 - m^2) . 
\nonumber \\ 
i \Delta_c^{(1)} (X; P) & = & 2 \pi \int d^{\, 3} q \, 
\frac{1}{\sqrt{E_+ E_-}} \nonumber \\ 
&& \times \left[ \langle a^\dagger ({\bf p}_+) a ({\bf 
p}_-) \rangle \delta (p_0 - (E_+ + E_-) / 2) 
e^{i[(E_+ - E_-) X_0 - {\bf q} \cdot {\bf X}]} \right. \nonumber \\ 
&& \mbox{\hspace*{3ex}} + \langle b^\dagger (- {\bf p}_+) b (- 
{\bf p}_-) \rangle \delta (p_0 + (E_+ + E_-) / 2) 
e^{i[(E_+ - E_-) X_0 + {\bf q} \cdot {\bf X}]} \nonumber \\ 
&& \mbox{\hspace*{3ex}} + \frac{1}{8} \langle a^\dagger 
({\bf p}_+) b^\dagger (- {\bf p}_-) \rangle \delta (p_0 - (E_+ - 
E_-) / 2) 
e^{i[(E_+ + E_-) X_0 - {\bf q} \cdot {\bf X}]} \nonumber \\ 
&& \mbox{\hspace*{3ex}} \left. + \frac{1}{8} \langle b (- 
{\bf p}_-) a ({\bf p}_+) \rangle \delta (p_0 - (E_+ - E_-) / 2) 
e^{- i[(E_+ + E_-) X_0 - {\bf q} \cdot {\bf X}]} \right] , 
\nonumber \\ 
\label{A1} 
\end{eqnarray} 
where ${\bf p}_\pm \equiv {\bf q} \pm {\bf q} / 2$ and $E_\pm \equiv 
E_{p_\pm}$. 

Let us first show that the contribution of the third and fourth 
terms of the bracketed quantities in (\ref{A1}) is negligibly small. 
From the setup in \S1, $\Delta_c^{(1)} (X; P)$ varies slowly in $X$, 
which means that $\langle a^\dagger ({\bf p}_+) b^\dagger (- {\bf 
p}_-) \rangle \sim \langle b (- {\bf p}_-) a ({\bf p}_+) \rangle 
\sim 0$ for $|{\bf q}| \gtrsim 1 / |{\bf L}|$ and $E_+ + E_- \gtrsim 
1 / L^0$, where $L^\mu$ is the size of a spacetime cell (see \S1). 
This can be realized only if $m \lesssim 1 / L^0$. However, as 
mentioned at the second footnote in \S4.3, in the case of $m 
\lesssim 1 / L^0$, the bare propagator should be replaced with the 
tadpole self-energy-part resummed propagator. The tadpole induces 
the mass of $O (\sqrt{\lambda {\cal T}})$ with ${\cal T}$ the 
typical scale(s) of the system, so that $E_+ + E_- >> 1 / L^0$. 

Let us turn to the first and second terms in (\ref{A1}). Same 
reasoning as above leads to $\langle a^\dagger ({\bf p}_+) a ({\bf 
p}_-) \rangle \sim \langle b^\dagger (- {\bf p}_+) b (- {\bf p}_-) 
\rangle \sim 0$ for $|{\bf q}| \gtrsim 1 / |{\bf L}|$. We shall 
consider the \lq\lq hard region'' $|P^\mu| > O (1 / L^\mu)$. From 
the first term, we pick out 
\begin{eqnarray*} 
&& \delta \left( p_0 - \frac{E_+ + E_-}{2} \right) e^{- i {\bf q} 
\cdot {\bf X}} \nonumber \\ 
& & \mbox{\hspace*{10ex}} = \left[ \delta (p_0 - E_p) - 
\frac{E^2_p q^2 - ({\bf p} \cdot {\bf q})^2}{8 E_p^3} \delta' (p_0 
- E_p) + \, ... \,  \right] e^{- i {\bf q} \cdot {\bf X}} \\ 
& & \mbox{\hspace*{10ex}} = \left[ \delta (p_0 - E_p) + \delta' 
(p_0 - E_p) \frac{E^2_p \nabla_{\bf x}^2 - ({\bf p} \cdot \nabla_{\bf 
x})^2}{8 E_p^3} + \, ... \, \right] e^{- i {\bf q} \cdot {\bf X}} . 
\end{eqnarray*} 
Then, to the gradient approximation, we have $\delta (p_0 - (E_+ + 
E_-) / 2) \simeq \delta (p_0 - E_p)$. Similarly, we have $E_+ E_- 
\simeq E_p^2$. The second term in (\ref{A1}) may be dealt with 
similarly. 

After all this, for $|P^\mu| > O (1 / L^\mu)$, we finally obtain 
\begin{eqnarray} 
i \Delta_c (X; P) & \simeq & 2 \pi \left[ \theta (p_0) \{1 + 2 
N_B^{(+)} (X; {\bf p}) \} \right. \nonumber \\ 
&& \left. + \theta (- p_0) \{1 + 2 N_B^{(-)} (X; - {\bf p}) \} 
\right] \delta (P^2 - m^2) , 
\label{Bpp} 
\end{eqnarray} 
where 
\begin{eqnarray} 
N_B^{(+)} (X; {\bf p}) & \equiv & \int d^{\, 3} q \langle 
a^\dagger ({\bf p} + {\bf q} / 2)  a ({\bf p} - {\bf q} / 2) 
\rangle \, \mbox{exp} \left[ i \left\{ \frac{{\bf p} \cdot {\bf 
q}}{E_p} X_0 - {\bf q} \cdot {\bf X} \right\} \right] , \nonumber \\ 
N_B^{(-)} (X; - {\bf p}) & \equiv & \int d^{\, 3} q \langle 
b^\dagger (- {\bf p} - {\bf q} / 2)  b (- {\bf p} + {\bf q} / 2) 
\rangle \, \mbox{exp} \left[ i \left\{ \frac{{\bf p} \cdot {\bf 
q}}{E_p} X_0 + {\bf q} \cdot {\bf X} \right\} \right] \nonumber \\ 
&& 
\label{di} 
\end{eqnarray} 
are the number densities of the quasiparticle ($+$) and the 
anti-quasiparticle ($-$). Comparison of (\ref{Bpp}) with 
(\ref{Deltac-1}) shows that 
\begin{eqnarray*} 
g_B^{(+)} (X; {\bf p}) & = & N_B^{(+)} (X; {\bf p}) , \\ 
g_B^{(-)} (X; {\bf p}) & = & - 1 - N_B^{(-)} (X; - {\bf p}) . 
\end{eqnarray*} 
For $\Delta_c (X; P)$ with $|P^\mu| \lesssim 1 / L^\mu$, one should 
use the fundamental formula (\ref{A1}). 

Incidentally, in the case where $\rho$ is diagonal in momentum 
space, $\langle a^\dagger ({\bf p}) a ({\bf q}) \rangle = \delta 
({\bf p} - {\bf q}) \tilde{n}_+ ({\bf p})$ and $\langle b^\dagger 
({\bf p}) b ({\bf q}) \rangle = \delta 
({\bf p} - {\bf q}) \tilde{n}_- ({\bf p})$, we obtain 
$N_B^{(\pm)} (X; {\bf p}) = \tilde{n}_\pm (\bf p)$, where 
$\tilde{n}_\pm$ are the number densities of the (anti)particle in 
vacuum theory. 
\setcounter{equation}{0}
\setcounter{section}{4}
\def\theequation{\mbox{\Alph{section}.\arabic{equation}}}
\section*{Appendix D Derivation of (\ref{iii}) and 
(\ref{bare-evo1})} 
\subsection*{D1 Derivation of (\ref{iii})} 
Let us start with computing the piece of (\ref{div}), which diverges 
in the narrow-width limit, $Im \Sigma_R \to - \epsilon (p_0) 0^+$. 
The relevant region (of integration) is where $Re (G_R^{- 1} (X; 
P)) = Re (G_A^{- 1} (X; P)) = P^2 - m^2 - Re \Sigma_R (X; P) \sim 
0$. We define \lq\lq on the mass-shell'' $p_0 = \pm \omega_\pm (X; 
\pm {\bf p})$ $(\equiv \pm \omega_\pm)$ through 
\begin{equation} 
[P^2 - m^2 - Re \Sigma_R (X; P))] \, \rule[-3mm]{.14mm}{8.5mm} 
\raisebox{-2.85mm}{\scriptsize{$\; p_0 = \pm \omega_\pm$}} = 0 , 
\label{Cyo} 
\end{equation} 
from which we obtain 
\begin{eqnarray} 
& & \pm Z_\pm^{- 1} \omega_\pm {\bf v}_\pm = {\bf p} + \frac{1}{2} 
\frac{\partial Re \Sigma_R}{\partial {\bf p}} \, 
\rule[-3mm]{.14mm}{8.5mm} \raisebox{-2.85mm}{\scriptsize{$\; p_0 
= \pm \omega_\pm$}} , \nonumber \\ 
& & \frac{\partial Re \Sigma_R}{\partial X^\mu} \, 
\rule[-3mm]{.14mm}{8.5mm} \raisebox{-2.85mm}{\scriptsize{$\; p_0 
= \pm \omega_\pm$}} = 2 Z_\pm^{- 1} \omega_\pm \frac{\partial 
\omega_\pm}{\partial X^\mu} . 
\label{kou} 
\end{eqnarray} 
Here ${\bf v}_\pm \equiv \pm \partial \omega_\pm / \partial {\bf p}$ 
are the velocity of the $\pm$ mode with momentum $\pm {\bf p}$ and 
\begin{equation} 
Z_\pm^{- 1} \equiv 1 \mp \frac{1}{2 \omega_\pm} \frac{\partial Re 
\Sigma_R}{\partial p_0} \rule[-3mm]{.14mm}{8.5mm} 
\raisebox{-2.85mm}{\scriptsize{$\; p_0 = \pm \omega_\pm$}} 
\label{Z} 
\end{equation} 
are wave-function renormalization factors. To extract the \lq\lq 
diverging'' piece, we can make following approximations; 
\begin{eqnarray} 
&& P^\mu \Sigma_{off} \sim \theta (p_0) \, (\omega_+, {\bf 
p})^\mu \, \Sigma_{off}^{(+)} + \theta (- p_0) \, (- \omega_-, 
{\bf p})^\mu \, \Sigma_{off}^{(-)} , \nonumber \\ 
&& \frac{- i}{2} \left[ G_R (X; P) - G_A (X; P) \right] \sim 
\sum_{\tau \pm} \frac{\theta (\tau p_0) \, Im 
\Sigma_R^{(\tau)}}{\left[ 2 \omega_\tau Z_\tau^{- 1} (p_0 - \tau 
\omega_\tau) \right]^2 + (Im \Sigma_R^{(\tau)})^2} , \nonumber \\ 
&& \label{koto} 
\end{eqnarray} 
where 
\begin{equation} 
\Sigma^{(\pm)}_{off} = \Sigma_{off} (X; p_0 = \pm 
\omega_\pm, {\bf p}), \;\;\; \Sigma^{(\pm)}_R = \Sigma_R (X; p_0 = 
\pm \omega_\pm, {\bf p}) . 
\label{hati} 
\end{equation} 
Using (\ref{koto}) in (\ref{div}), we obtain 
\begin{equation} 
\langle \delta j^\mu (X) \rangle \simeq \frac{1}{2} \int \frac{d^{\, 
3} p}{(2 \pi)^3} \left[ \hat{\cal P}_+^\mu Z_+ \frac{- i 
\Sigma_{off}^{(+)}}{|Im \Sigma_R^{(+)}|} - \hat{\cal P}_-^\mu Z_- 
\frac{- i \Sigma_{off}^{(-)}}{|Im \Sigma_R^{(-)}|} \right] , 
\label{ii} 
\end{equation} 
where $\hat{\cal P}_\pm^\mu \equiv (1, \pm {\bf p} / \omega_\pm)$. 
In deriving (\ref{ii}), we have assumed that $|Im \Sigma_R^{(\pm)}| 
<< \omega_\pm$, under which (\ref{koto}) is a good approximation. 
Referring to (\ref{iik}), we can extract from (\ref{ii}) the 
contribution to the physical-number density: 
\[ 
\delta N^{(\pm)} (X; \pm \omega_\pm, \pm \hat{\bf p}) = 
\frac{Z_\pm}{2} \frac{- i \Sigma_{off}^{(\pm)}}{|Im 
\Sigma_R^{(\pm)}|} . 
\] 
\subsection*{D2 Derivation of (\ref{bare-evo1})} 
We note that 
$\Gamma_\pm^{(p)} \tau_\pm / 2$ in (\ref{iii}) is the change in the 
\lq\lq physical''-number density, during the time interval $\tau_\pm 
/ 2$, due the the net production rate. In \S1, we have introduced 
the spacetime cells, whose size is $L^\mu$ $(\mu = 0, 1, 2, 3)$. It 
is quite natural to take $\tau_\pm = L_0$, where $L_0$ is the size 
of the time direction of the spacetime cell (including the spacetime 
point $X^\mu$ in (\ref{iii})). [Strictly speaking, in general, 
$\tau_+ \neq \tau_-$. However, in the present crude argument, we 
ignore this difference.] It is interesting to note that 
$\Gamma_\pm^{(p)} \tau_\pm / 2$ in (\ref{iii}) is {\em half} of the 
net production probability, due to the reactions, during the time 
interval $\tau_\pm = L_0$. [In this respect, cf. 
Ref.~21).] 

Noticing that we are concerned about the quasiparticle mode with 
momentum ${\bf p}$ [anti-quasiparticle mode with momentum $- {\bf 
p}$], we see that (\ref{iii}) leads to the relations, 
\begin{eqnarray} 
&& \delta N_\pm (X_0 + L_0, {\bf X} + {\bf v}_\pm L_0; \pm 
\omega_\pm, \pm \hat{\bf p}) - \delta N_\pm (X_0, {\bf X}; \pm 
\omega_\pm, \pm \hat{\bf p}) \nonumber \\ 
&& \mbox{\hspace*{10ex}} \simeq L_0 Z_\pm \frac{- i \Sigma_{off} 
(X; \pm \omega_\pm, {\bf p})}{2 \omega_\pm} , 
\label{ara} 
\end{eqnarray} 
from which we obtain 
\begin{equation} 
\frac{\partial \delta N_\pm (X; \pm \omega_\pm, \pm \hat{\bf 
p})}{\partial X_0} + {\bf v}_\pm \cdot \frac{\partial \delta 
N_\pm}{\partial {\bf X}} \simeq  Z_\pm \left[ \Gamma_\pm^{(p)} + 
\frac{\left\{ f_B, \; Re \Sigma_R \right\}}{2 \omega_\pm} 
\right]_{p_0 = \pm \omega_\pm} . 
\label{bare-evo} 
\end{equation} 
Using (\ref{kou}) with (\ref{Z}), we get 
\begin{eqnarray*} 
\left\{ f_B, \; Re \Sigma_R \right\} \, \rule[-3mm]{.14mm}{8.5mm} 
\raisebox{-2.85mm}{\scriptsize{$\; p_0 = \pm \omega_\pm$}} & = & 
- 2 Z_\pm^{- 1} \omega_\pm \left[ \left( {\bf v}_\pm \mp \frac{\bf 
p}{\omega_\pm} \right) \frac{\partial N_B^{(\pm)} (X; \pm 
\omega_\pm, \pm \hat{\bf p})}{\partial {\bf X}} \right. \nonumber \\ 
&& \left. \pm \frac{\partial \omega_\pm}{\partial X^\mu} 
\frac{\partial N_B^{(\pm)} (X; \pm p_0, \pm \hat{\bf p})}{\partial 
P_\mu} \, \rule[-3mm]{.14mm}{8.5mm} 
\raisebox{-2.85mm}{\scriptsize{$\; p_0 = \pm \omega_\pm$}} \right] 
. 
\end{eqnarray*} 
where use has been made of (\ref{const}) and (\ref{p0f}). 
Substituting this into (\ref{bare-evo}) and adding (\ref{const}), we 
finally obtain 
\[ 
\frac{\partial N^{(\pm)}}{\partial X_0} + {\bf v}_\pm \cdot 
\nabla_{\bf X} N^{(\pm)} \simeq \left[ Z_\pm \Gamma_\pm^{(p)} \mp 
\frac{\partial \omega_\pm}{\partial X_\mu} \frac{\partial 
N^{(\pm)}_B (X; \pm p_0, \pm \hat{\bf p})}{\partial P^\mu} 
\right]_{p_0 = \pm \omega_\pm} , 
\] 
where $N^{(\pm)} = N_B^{(\pm)} (X; \pm \omega_\pm, \pm \hat{\bf p}) 
+ \delta N_\pm (X; \pm \omega_\pm, \pm \hat{\bf p})$. 

\end{document}